\documentstyle[aps,prb,epsf,twocolumn]{revtex}

\begin{document}
\twocolumn[
\begin{center}
\vspace{0mm}
{\large{\bf Gauge Invariance and Hall Terms in 
the Quasiclassical Equations of Superconductivity}}

\vspace{4mm}
{Takafumi Kita}
\par

\vspace{1mm}
{\small\em Physikalisches Institute,
Universit\"at Bayreuth,
95440 Bayreuth, Germany\\
and\\
Division of Physics, Hokkaido University,
Sapporo 060-0810, Japan}
\par

({\today})
\vspace{3mm}

{\small \parbox{142mm}{\hspace*{5mm}
This paper presents a careful derivation of the
quasiclassical equations of superconductivity so that
a manifest gauge invariance is retained
with respect to the space-time arguments of the
quasiclassical Green's function $\hat{g}$.
The terms responsible for the Hall effect
naturally appear from the derivation.
The equations are applicable to clean as well as dirty superconductors
for an arbitrary external frequency
much smaller than the Fermi energy.
Thus, they will form a basis toward a complete microscopic understanding 
of the Hall effect in type-II superconductors.
}
}
\end{center}

\vspace{8mm}
]

\section{Introduction}
\label{sec:intro}

The Hall effect in the vortex state of type-II superconductors remains a matter of 
controversy after decades of intensive investigations.
The early phenomenological theories of Bardeen and Stephen\cite{BS65} and 
Nozi\`eres and Vinen\cite{NV66} fail to account for the sign change of the
Hall conductivity found in the vortex state of a wide variety
of materials.\cite{Hagen93,Nagaoka98}
Also, a debate still continues about 
the forces acting on a single moving vortex.\cite{Sonin97}
This state of affairs may be attributed partly to a lack
of the established tractable microscopic equations
with which one could test the validity of various phenomenological models
numerically.
Especially, the standard quasiclassical equations of superconductivity, 
i.e.\ one of the most powerful methods for nonequilibrium
superfluids and superconductors,\cite{Rainer83}
have been known to be unable to describe the phenomenon.

Efforts have been made recently to include terms
responsible for the Hall effect in the quasiclassical equations.
Larkin and Ovchinnikov\cite{LO95} incorporated higher-order effects
arising from the particle-hole asymmetry near the Fermi level,
but the main terms corresponding the normal-state Hall effect are
still missing in their equations.
Kopnin\cite{Kopnin94} obtained kinetic equations with the desired Hall terms
and used them to discuss the flux-flow Hall effect.\cite{Kopnin95}
However, their applicability is limited to clean superconductors 
with slow time variations,
due to his transformation to Green's functions 
[see Eq.\ (\ref{Idef-Kopnin}) below]
which may not be suitable for deriving the equations 
for high-frequency disturbances.
More recently, Houghton and Vekhter\cite{HV98} also reported an extension,
but there seem to be a couple of unsatisfactory points.
First, the obtained equations do not carry a manifest gauge 
invariance with respect to the space-time arguments 
of the quasiclassical Green's function $\hat{g}$.
The second point lies in their derivation process:
They first define the local one-particle energy 
$\xi\!\equiv\!\xi({\bf p}\!-\!\frac{e}{c}{\bf A})$
which depends on the position ${\bf R}$ as well as the momentum ${\bf p}$
through the spatial dependence of the vector potential ${\bf A}\!=\!{\bf A}({\bf R})$.
Whereas they expand this $\xi$ with respect to ${\bf A}$ to get the linear field dependence
of the equation, the solution $\hat{g}$ to the equation
is formally defined by the integral of the full Green's function $\hat{G}$
over the unexpanded $\xi\!=\!\xi({\bf p}\!-\!\frac{e}{c}{\bf A})$.
The validity of this procedure is not entirely clear.

With these observations, we here present a careful and 
more straightforward derivation of the quasiclassical equations
which can fully describe the Hall effect of the vortex states
and form a firm basis for any detailed numerical studies.
A key ingredient lies in the introduction of a new transformed
Green's function [see Eq.\ (\ref{HatGbar-R}) below] whose gauge change 
can solely be expressed with respect to the 
slowly varying space-time coordinate;
they are different from those used by Kopnin\cite{Kopnin94}
and enable us to obtain completely gauge-invariant quasiclassical equations.
The idea goes back to the original work of Gor'kov\cite{Gor'kov59}
and was extensively used by Eilenberger\cite{Eilenberger66}
prior to his derivation
of the quasiclassical equations.\cite{Eilenberger68}
On the other hand, Larkin and Ovchinnikov\cite{LO68}
presented a more compact derivation
of the quasiclassical equations
with the left-right subtraction trick
without recourse to the transformed Green's function.
Those two approaches certainly provide the same equations 
at the lowest order of the approximation.
It turns out, however, that the advantages of the two approaches
have to be combined to proceed further
with the gauge-invariance at every order of the approximation
for a systematic derivation of the Hall terms.
Indeed, the quasiclassical equations will be obtained here by applying 
the left-right subtraction trick 
to the Dyson-Gor'kov equation for the new transformed Green's function.

This paper is organized as follows. 
Section \ref{sec:Green} 
writes down the Dyson-Gor'kov equation for the 
conventional retarded Nambu Green's function,
followed by an introduction of
a new Green's function 
whose gauge change can be expressed
only with respect to the center-of-mass coordinate.
Section \ref{sec:space-time} transforms the space-time derivatives
of the Dyson-Gor'kov equation into an expression
of using the new Green's function.
Section \ref{sec:self-energy} performs a similar
transformation to the self-energy part
of the Dyson-Gor'kov equation.
Section \ref{sec:quasi} collects the results of Secs.\ \ref{sec:space-time}
and \ref{sec:self-energy} to write down the Dyson-Gor'kov equation
for the transformed Green's function, and subsequently
derives the retarded quasiclassical equations
by the left-right subtraction trick.
Section \ref{sec:Nambu-Keldysh} presents the
extension to the advanced and the Keldysh parts.
Section \ref{sec:conclusion} concludes the paper with several remarks.

We put $\hbar\!=\!1$ throughout, and denote the light velocity, 
the electron charge, and the electron bare mass
by $c$, $e(<0)$, and $m$, respectively.

\section{Green's functions}
\label{sec:Green}

We will consider the elements of the Nambu-Keldysh matrix\cite{Rainer83} separately
as it turns out to be more convenient than handling the matrix itself.
We first focus on the retarded part to describe the derivation,
and then carry out the extension to the advanced and the Keldysh parts.
We hence drop the conventional superscript $^{\rm R}$ signifying
``retarded'' while the distinctions are unnecessary.

Let us define a couple of retarded Green's functions by
\begin{equation}
G_{\alpha\beta}(1,2)\equiv -i\theta(t_{1}\!-\!t_{2})\langle 
\{\psi_{\alpha}(1),\psi_{\beta}^{\dagger} (2)\}\rangle \, , \\
\label{Gdef}
\end{equation}
\begin{equation}
F_{\alpha\beta}(1,2)  \equiv -i\theta(t_{1}\!-\!t_{2})\langle 
\{\psi_{\alpha}(1),\psi_{\beta} (2)\}\rangle \, ,
\label{Fdef}
\end{equation}
where $1\!\equiv\!{\bf r}_{1}t_{1}$ specifies the space-time coordinate,
$\alpha\beta$ are the spin indices, and 
$\{A,B\}\!\equiv \! AB\!+\! BA$.
To suppress the spin indices we introduce
$2\!\times\!2$ matrices $\underline{G}$ and $\underline{F}$ by
$(\underline{G})_{\alpha\beta}\!=\! G_{\alpha\beta}$ and
$(\underline{F})_{\alpha\beta}\!=\! F_{\alpha\beta}$.
Using $\underline{G}$ and $\underline{F}$,
we next define a $4\times 4$ Nambu matrix by\cite{comment1}
\begin{eqnarray}
&&\hat{G}(1,2) \equiv 
\left[\!
\begin{array}{cc}
\vspace{1mm}
\underline{G}(1,2) & \underline{F}(1,2)
\\
-\underline{F}^{*}(1,2) & -\underline{G}^{*}(1,2)
\end{array}
\!\right] ,
\label{HatG-R}
\end{eqnarray}
and the corresponding self-energy matrix by
\begin{eqnarray}
&&\hat{\Sigma}(1,2) \equiv 
\left[\!
\begin{array}{cc}
\vspace{1mm}
\underline{\Sigma}(1,2) & \underline{\Delta}(1,2)
\\
-\underline{\Delta}^{\! *\!}(1,2) & -\underline{\Sigma}^{*\!}(1,2)
\end{array}
\!\right]  .
\label{HatSig-R}
\end{eqnarray}
They are changed for
$\psi_{\alpha}(1)\rightarrow
\psi_{\alpha}(1) \exp[i\frac{e}{c}\chi(1)]$
as
$\hat{G}(1,2) \rightarrow$
$\exp[i\frac{e}{c}\chi(1)\hat{\tau}_{3}]
\hat{G}(1,2)\exp[-i\frac{e}{c}\chi(2)\hat{\tau}_{3}]$,
where $\hat{\tau}_{3}$ is defined by
\begin{eqnarray}
\hat{\tau}_{3} \equiv 
\left[\begin{array}{rr}
\underline{1} & \underline{0} \\
\underline{0} & -\underline{1}
\end{array}\right] ,
\end{eqnarray}
with $\underline{1}$ and $\underline{0}$ denoting
the $2\times 2$ unit and zero matrices, respectively.
They satisfy the Dyson-Gor'kov equation:
\begin{eqnarray}
&&\left(i\frac{\partial}{\partial t_{1}}\!-\! e\Phi_{1}\hat{\tau}_{3}\right)\hat{G}(1,2)
-\left[
\begin{array}{cc}
\underline{\cal H}_{1} & \underline{0}
\\
\underline{0} & -\underline{\cal H}^{*}_{1}
\end{array} 
\right]
\hat{G}(1,2)
\nonumber \\
&&-\!\int\hat{\Sigma}(1,3)\hat{G}(3,2)
d3 = \delta(1-2) \hat{1}\, .
\label{Dyson-R}
\end{eqnarray}
Here $\Phi$ is the scalar potential,
$\hat{1}$ is the $4\times 4$ unit matrix,
and $\underline{\cal H}_{1}$ is defined by
\begin{eqnarray}
\underline{\cal H}_{1}\equiv
\left[\frac{1}{2m}\left(-i\frac{\partial}{\partial {\bf r}_{1}}
-\frac{e}{c}{\bf A}_{1}\right)^{\!\! 2}
-\mu\right] \underline{1}\, ,
\label{K}
\end{eqnarray}
with $\mu$ the chemical potential
and $\bf A$ the vector potential.

We now introduce the key quantities,
i.e.\ a couple of new Nambu matrices, 
by using a nonlocal gauge transformation
as
\begin{eqnarray}
\hat{\bar{G}}(1,2)  
\equiv \exp[iI(\vec{R},\vec{r}_{1})\hat{\tau}_{3}]
\hat{G}(1,2)
\exp[-iI(\vec{R},\vec{r}_{2})\hat{\tau}_{3}] \, ,
\label{HatGbar-R}
\end{eqnarray}
\begin{eqnarray}
\hat{\bar{\Sigma}}(1,2)  
\equiv \exp[iI(\vec{R},\vec{r}_{1})\hat{\tau}_{3}]
\hat{\Sigma}(1,2)
\exp[-iI(\vec{R},\vec{r}_{2})\hat{\tau}_{3}]
\, .
\label{HatSbar-R}
\end{eqnarray}
Here $\vec{r}_{1}\!\equiv\!(ct_{1},{\bf r}_{1})$ 
is the four vector,\cite{LL} $\vec{R}\!\equiv\!\vec{R}_{12}
\!\equiv\!\frac{1}{2}(\vec{r}_{1}+\vec{r}_{2})$,
and $I$ is defined by
\begin{eqnarray}
I(\vec{R},\vec{r}_{1})\equiv -\frac{e}{c}\int_{\vec{r}_{1}}^{\vec{R}}\!
\vec{A}(\vec{s})\cdot d\vec{s} \, ,
\label{Idef}
\end{eqnarray}
where $\vec{A}\!\equiv\!(\Phi,-{\bf A})$ denotes the covariant electromagnetic potential
and $d\vec{s}$ is taken along the straight line.

The quantities $\hat{\bar{G}}$ and $\hat{\bar{\Sigma}}$ defined above
have a desired property that
only the center-of-mass coordinate $\vec{R}$ is relevant
in the gauge transformation $\psi_{\alpha}(1)\rightarrow
\psi_{\alpha}(1) \exp[i\frac{e}{c}\chi(1)]$.
Indeed, $\hat{\bar{G}}$ is changed as
$\hat{\bar{G}}(1,2) \rightarrow\exp[i\frac{e}{c}\chi(\vec{R})\hat{\tau}_{3}]
\hat{\bar{G}}(1,2)\exp[-i\frac{e}{c}\chi(\vec{R})\hat{\tau}_{3}]$.
Proceeding with $\hat{\bar{G}}$ and $\hat{\bar{\Sigma}}$,
we are led to the equations with a manifest gauge invariance
with respect to $\vec{R}$, as seen below.
Indeed, Levanda and Fleurov\cite{LF94} has successfully derived
normal-state kinetic equations with a manifest gauge invariance
by using the component $\underline{\bar{G}}$ of Eq.\ (\ref{HatGbar-R}).

It is worth pointing out the difference of the above $\hat{\bar G}$ from
that used by Kopnin.
As mentioned in Introduction, Kopnin\cite{Kopnin94} has derived kinetic equations 
based on a couple of transformed Green's functions
similar to Eq.\ (\ref{HatGbar-R}).
However, his phase factor is different from Eq.\ (\ref{Idef}) as
\begin{eqnarray}
I_{\rm K}(\vec{R},\vec{r}_{1})\equiv -e\int_{t_{1}}^{T}\!
\Phi({\bf r}_{1}t')\, dt'+ \frac{e}{c}\int_{{\bf r}_{1}}^{{\bf R}}\!
{\bf A}({\bf r}'t_{1})\cdot d{\bf r}'  ,
\label{Idef-Kopnin}
\end{eqnarray}
where $d{\bf r}'$ is along the straight line;
see Eq.\ (18) of Ref.\ \onlinecite{Kopnin94}.
The present $\hat{\bar G}$ may be more advantageous,
since its gauge transformation property is expressible only with respect to $\vec R$.
Indeed, it will enable us a more systematic and comprehensive derivation of the Hall terms.

It is convenient for later purposes to introduce the two functions:
\begin{equation}
{\cal E}_{1}(u) \equiv \int_{0}^{1}\! d\eta\, {\rm e}^{\eta u}
=\frac{{\rm e}^{u}-1}{u} \, ,
\label{CalE1}
\end{equation}
\begin{equation}
{\cal E}_{2}(u) \equiv \int_{0}^{1}\!d\eta\int_{0}^{\eta}\!d\zeta \, {\rm e}^{\zeta u}  
=\frac{{\rm e}^{u}-1-u}{u^{2}} \, .
\label{CalE2}
\end{equation}
Then Eq.\ (\ref{Idef}) can also be written as
\begin{eqnarray}
I(\vec{R},\vec{r}_{1}) = \frac{e}{c}\,
{\cal E}_{1}\!\!\left(\frac{\vec{r}}{2}\!\cdot\!\frac{\partial}{\partial \vec{R}}
\right)\frac{\vec{r}}{2}\!\cdot\!\vec{A}(\vec{R}) \, ,
\label{Idef2}
\end{eqnarray}
with $\vec{r}\!\equiv\!\vec{r}_{12}\!\equiv\!\vec{r}_{1}\!-\!\vec{r}_{2}$.

\section{Space-time derivatives}
\label{sec:space-time}

Let us rewrite the first two terms on the left-hand side of 
Eq.\ (\ref{Dyson-R})
with respect to $\hat{\bar{G}}$ of Eq.\ (\ref{HatGbar-R}).
The following identities are useful for this purpose ($j\!=\!0,1,2,3$):
\begin{eqnarray}
&&\frac{\partial}{\partial r_{1j}}I(\vec{R},\vec{r}_{1})
= \frac{e}{c}A_{j}(\vec{r}_{1})\!-\!\frac{e}{2c}A_{j}(\vec{R})
\nonumber \\
&&-\frac{e}{4c}\, [2{\cal E}_{1}(\hat{u})\!-\!{\cal E}_{2}(\hat{u})]\,
r_{k}\!\left[\frac{\partial \vec{A}_{j}(\vec{R})}{\partial R_{k}}\!
-\!\frac{\partial \vec{A}_{k}(\vec{R})}{\partial R_{j}}\right]  ,
\label{dI1}
\end{eqnarray}
\begin{eqnarray}
&&\frac{\partial}{\partial r_{1j}}I(\vec{R},\vec{r}_{2})
= -\frac{e}{2c}A_{j}(\vec{R})
\nonumber \\
&&+\frac{e}{4c}{\cal E}_{2}(-\hat{u})\,
r_{k}\!\left[\frac{\partial \vec{A}_{j}(\vec{R})}{\partial R_{k}}\!
-\!\frac{\partial \vec{A}_{k}(\vec{R})}{\partial R_{j}}\right]  ,
\label{dI2}
\end{eqnarray}
where $\hat{u}\equiv\frac{\vec{r}}{2}\cdot\frac{\partial}{\partial \vec{R}}$,
and summations over the repeated index $k$ $(=\!0,1,2,3)$ are implied.
Using the results, the gauge-invariant time and space derivatives 
of $\underline{G}$ are transformed as
\begin{eqnarray}
&&{\rm e}^{iI(\vec{R},\vec{r}_{1})-iI(\vec{R},\vec{r}_{2})}
\left[i\frac{\partial}{\partial t_{1}}
\!-\!e\Phi(\vec{r}_{1}) \right]\! \underline{G}(1,2)
\nonumber \\
= && \biggl\{i\frac{\partial}{\partial t}+\frac{i}{2}\frac{\partial}{\partial T}
+\frac{e}{4}\bigl[\, 
2{\cal E}_{1}\!\left({\textstyle \frac{t}{2}\frac{\partial}{\partial T}}
\right)-{\cal E}_{2}\!\left({\textstyle \frac{t}{2}\frac{\partial}{\partial T}}
\right)
\nonumber \\
&&
+{\cal E}_{2}\!\left(-{\textstyle \frac{t}{2}\frac{\partial}{\partial T}}
\right)\,\bigr]\,
{\bf E}(\vec{R})\!\cdot\!{\bf r} \biggr\} \underline{\bar{G}}(1,2)\, ,
\label{dGdt}
\end{eqnarray}
and
\begin{eqnarray}
&&{\rm e}^{iI(\vec{R},\vec{r}_{1})-iI(\vec{R},\vec{r}_{2})}
\left[-i\frac{\partial}{\partial {\bf r}_{1}}
\!-\!\frac{e}{c}{\bf A}(\vec{r}_{1}) \right]\! \underline{G}(1,2)
\nonumber \\
 = && \biggl\{-i\frac{\partial}{\partial {\bf r}}-\frac{i}{2}\frac{\partial}{\partial {\bf R}}
-\frac{e}{4}\bigl[\, 
2{\cal E}_{1}\!\left({\textstyle \frac{t}{2}\frac{\partial}{\partial T}}
\right)-{\cal E}_{2}\!\left({\textstyle \frac{t}{2}\frac{\partial}{\partial T}}
\right)
\nonumber \\
&&
+{\cal E}_{2}\!\left(-{\textstyle \frac{t}{2}\frac{\partial}{\partial T}}
\right)\,\bigr]\!\left[ \, \frac{1}{c}{\bf h}(\vec{R})\!\times\!{\bf r}
\!-\!{\bf E}(\vec{R})\hspace{0.3mm}t \, \right]
\biggr\} \underline{\bar{G}}(1,2)\, ,
\label{dGdr}
\end{eqnarray}
where we have neglected spatial derivatives of both the electric field ${\bf E}$ and 
the magnetic field ${\bf h}$.
Similarly, we have
\begin{eqnarray}
&&{\rm e}^{iI(\vec{R},\vec{r}_{1})+iI(\vec{R},\vec{r}_{2})}
\left[i\frac{\partial}{\partial t_{1}}
\!-\!e\Phi(\vec{r}_{1}) \right]\! \underline{F}(1,2)
\nonumber \\
= && \biggl\{i\frac{\partial}{\partial t}+\frac{i}{2}\frac{\partial}{\partial T}
-e\Phi(\vec{R})+\frac{e}{4}\bigl[\, 
2{\cal E}_{1}\!\left({\textstyle \frac{t}{2}\frac{\partial}{\partial T}}
\right)-{\cal E}_{2}\!\left({\textstyle \frac{t}{2}\frac{\partial}{\partial T}}
\right)
\nonumber \\
&&
-{\cal E}_{2}\!\left(-{\textstyle \frac{t}{2}\frac{\partial}{\partial T}}
\right)\,\bigr]\,
{\bf E}(\vec{R})\!\cdot\!{\bf r} \biggr\} \underline{\bar{F}}(1,2)\, ,
\label{dFdt}
\end{eqnarray}
and
\begin{eqnarray}
&&{\rm e}^{iI(\vec{R},\vec{r}_{1})+iI(\vec{R},\vec{r}_{2})}
\left[-i\frac{\partial}{\partial {\bf r}_{1}}
\!-\!\frac{e}{c}{\bf A}(\vec{r}_{1}) \right]\! \underline{F}(1,2)
\nonumber \\
 = && \biggl\{-i\frac{\partial}{\partial {\bf r}}-\frac{i}{2}\frac{\partial}{\partial {\bf R}}
-\frac{e}{c}{\bf A}(\vec{R})
-\frac{e}{4}\bigl[\, 
2{\cal E}_{1}\!\left({\textstyle \frac{t}{2}\frac{\partial}{\partial T}}
\right)-{\cal E}_{2}\!\left({\textstyle \frac{t}{2}\frac{\partial}{\partial T}}
\right)
\nonumber \\
&&
-{\cal E}_{2}\!\left(-{\textstyle \frac{t}{2}\frac{\partial}{\partial T}}
\right)\,\bigr]\!\left[ \, \frac{1}{c}{\bf h}(\vec{R})\!\times\!{\bf r}
\!-\!{\bf E}(\vec{R})\hspace{0.3mm}t \, \right]
\biggr\} \underline{\bar{F}}(1,2)\, .
\label{dFdr}
\end{eqnarray}
Notice the differences of $\pm{\cal E}_{2}
(-{\textstyle \frac{t}{2}\frac{\partial}{\partial T}}
)$
between Eqs.\ (\ref{dGdt}) and (\ref{dFdt}), and also
between Eqs.\ (\ref{dGdr}) and (\ref{dFdr}).

We now introduce the Fourier transform of $\hat{\bar{G}}$ by
\begin{eqnarray}
&&\hat{G}({\bf p}\varepsilon ,{\bf R}T) \equiv \int
\hat{\bar{G}}(1,2) 
\,{\rm e}^{-i({\bf p}\cdot{{\bf r}}-\varepsilon t)} 
\, d{\bf r}dt
\nonumber \\
&& \equiv\left[\!
\begin{array}{cc}
\vspace{1mm}
\underline{G}({\bf p}\varepsilon ,{\bf R}T) & 
\underline{F}({\bf p}\varepsilon ,{\bf R}T)
\\
-\underline{F}^{*\!}(-{\bf p}\!-\!\varepsilon ,{\bf R}T) & 
-\underline{G}^{*\!}(-{\bf p}\!-\!\varepsilon ,{\bf R}T)
\end{array}
\!\right] ,
\label{HatG}
\end{eqnarray}
where the arguments $\varepsilon$ and $-\varepsilon$ both have an infinitesimal
positive imaginary part.
We also define the gauge-invariant derivatives
$\partial_{T}$ and
$\mbox{\boldmath$\partial$}_{{\bf R}}$ by
\begin{eqnarray}
\partial_{T}\equiv\left\{
\begin{array}{ll}
\vspace{1mm}
\frac{\partial}{\partial T} & 
: \mbox{on $\underline{G}$, $\underline{G}^{*}$, 
$\underline{\Sigma}$, $\underline{\Sigma}^{*}$, ${\bf E}$, ${\bf h}$}
\\
\vspace{1mm}
\frac{\partial}{\partial T}\!+\!2ie\Phi(\vec{R}) & 
: \mbox{on $\underline{F}$, $\underline{\Delta}$}
\\
\frac{\partial}{\partial T} \!-\!2ie\Phi(\vec{R}) & 
: \mbox{on $\underline{F}^{*}$, $\underline{\Delta}^{*}$}
\end{array}
\right. ,
\label{partialT}
\end{eqnarray}
\begin{eqnarray}
\mbox{\boldmath$\partial$}_{\bf R}\equiv\left\{
\begin{array}{ll}
\vspace{1mm}
\frac{\partial}{\partial {\bf R}} & 
: \mbox{on $\underline{G}$, $\underline{G}^{*}$, 
$\underline{\Sigma}$, $\underline{\Sigma}^{*}$, ${\bf E}$, ${\bf h}$}
\\
\vspace{1mm}
\frac{\partial}{\partial {\bf R}}\!-\!i\frac{2e}{c}{\bf A}(\vec{R}) & 
: \mbox{on $\underline{F}$, $\underline{\Delta}$}
\\
\frac{\partial}{\partial {\bf R}} \!+\! i\frac{2e}{c}{\bf A}(\vec{R}) & 
: \mbox{on $\underline{F}^{*}$, 
$\underline{\Delta}^{*}$}
\end{array}
\right. .
\label{partialR}
\end{eqnarray}
Then using Eqs.\ (\ref{dGdt})-(\ref{dFdr}), we can write
the first two terms on the left-hand side of 
Eq.\ (\ref{Dyson-R}) with respect to Eqs.\ 
(\ref{HatG})-(\ref{partialR}).
We finally neglect terms second-order in 
$\mbox{\boldmath$\partial$}_{{\bf R}}$,
${\bf E}$, ${\bf h}$, and 
$({\bf p}-{\bf p}_{\rm F})^{2}$ with ${\bf p}_{\rm F}$
the Fermi momentum,
since they are smaller than
the first-order ones by an order of magnitude 
in ``small'',\cite{Rainer83} i.e.\ $(p_{\rm F}\xi_{0})^{-1}$ etc,
with $\xi_{0}$ the coherence length.
With these procedures,
the first two terms on the left-hand side of 
Eq.\ (\ref{Dyson-R}) are Fourier-transformed into
\begin{eqnarray}
&&\left(\varepsilon+\frac{i}{2}\partial_{T}\right)\hat{G}
-\left[\frac{p_{\rm F}^2}{2m}-\mu
+ \frac{{\bf p}_{\rm F}}{m}\cdot({\bf p}-{\bf p}_{\rm F})
\right]
\hat{\tau}_{3}\hat{G}
\nonumber \\
&&+ \frac{i}{2}\,\frac{{\bf p}_{\rm F}}{m}\cdot
\mbox{\boldmath$\partial$}_{\bf R}\hat{\tau}_{3}\hat{G}
\nonumber \\
&&+\frac{i}{4}e\bigl[\, 
2{\cal E}_{1}\!\left({\textstyle -\frac{i}{2}\partial_{T}\partial_{{\varepsilon}}}
\right)-{\cal E}_{2}\!\left(-{\textstyle \frac{i}{2}\partial_{T}\partial_{{\varepsilon}}}
\right)\bigr]
{\bf E}\cdot \mbox{\boldmath$\partial$}_{{\bf p}}\hat{\tau}_{3}\hat{G}
\nonumber \\
&&+\frac{i}{4}e\,
{\cal E}_{2}\!\left({\textstyle \frac{i}{2}\partial_{T}\partial_{{\varepsilon}}}
\right)
{\bf E}\cdot \mbox{\boldmath$\partial$}_{{\bf p}}\hat{G}\hat{\tau}_{3}
\nonumber \\
&&+\frac{i}{4}e \bigl[\, 
2{\cal E}_{1}\!\left({\textstyle -\frac{i}{2}\partial_{T}\partial_{{\varepsilon}}}
\right)-{\cal E}_{2}\!\left(-{\textstyle \frac{i}{2}\partial_{T}\partial_{{\varepsilon}}}
\right)\bigr]
\nonumber \\
&&\hspace{30mm}\times
\frac{{\bf p}_{\rm F}}{m}\cdot 
\left(\frac{1}{c}{\bf h}\times\mbox{\boldmath$\partial$}_{{\bf p}}
+{\bf E}\, \partial_{\varepsilon}\right)\hat{G}
\nonumber \\
&&+\frac{i}{4}e\, {\cal E}_{2}\!\left({\textstyle \frac{i}{2}\partial_{T}\partial_{{\varepsilon}}}
\right)
\frac{{\bf p}_{\rm F}}{m}\cdot 
\left(\frac{1}{c}{\bf h}\times\mbox{\boldmath$\partial$}_{{\bf p}}
+{\bf E}\, \partial_{\varepsilon}\right)\hat{\tau}_{3}\hat{G}\hat{\tau}_{3}
\, ,
\label{DysonL-K}
\end{eqnarray}
where
$\mbox{\boldmath$\partial$}_{\bf p}\equiv\frac{\partial}{\partial {\bf p}}$,
$\partial_{\varepsilon}\equiv\frac{\partial}{\partial \varepsilon}$,
and ${\cal E}_{1}$ and ${\cal E}_{2}$  are defined by
Eqs.\ (\ref{CalE1}) and (\ref{CalE2}), respectively.
Here one should keep in mind that $\partial_{T}$ in 
${\cal E}_{1}$ and ${\cal E}_{2}$
operates only on ${\bf E}$ and ${\bf h}$.

With no time dependence in ${\bf E}$ and ${\bf h}$,
${\cal E}_{1}\rightarrow 1$ and ${\cal E}_{2}\rightarrow \frac{1}{2}$
so that the last four terms in Eq.\ (\ref{DysonL-K}) reduce to
\begin{eqnarray}
&&\frac{i}{8}e\,
{\bf E}\cdot \mbox{\boldmath$\partial$}_{{\bf p}}
(3\hat{\tau}_{3}\hat{G}+\hat{G}\hat{\tau}_{3})
\nonumber \\
+&&\frac{i}{8}e\,
\frac{{\bf p}_{\rm F}}{m}\cdot 
\left(\frac{1}{c}{\bf h}\times\mbox{\boldmath$\partial$}_{{\bf p}}
+{\bf E}\, \partial_{\varepsilon}\right)
(3\hat{G}+\hat{\tau}_{3}\hat{G}\hat{\tau}_{3})
\, .
\label{DysonL-K-St}
\end{eqnarray}
These terms are absent in the conventional derivations,
which are certainly responsible for
the Hall effect of both the normal and the vortex states.

Before closing the section,
we compare the above result with
the corresponding one derived by Kopnin.\cite{Kopnin94}
Due to the difference between Eqs.\ (\ref{Idef}) and (\ref{Idef-Kopnin}),
he obtained instead of Eq.\ (\ref{DysonL-K-St}) the expression:
\begin{eqnarray}
\frac{i}{2}e \left[{\bf E}\cdot \mbox{\boldmath$\partial$}_{{\bf p}}
+
\frac{{\bf p}_{\rm F}}{m}\cdot 
\left(\frac{1}{c}{\bf h}\times\mbox{\boldmath$\partial$}_{{\bf p}}
+{\bf E}\, \partial_{\varepsilon}\right)\right]
\hat{G}
\, .
\label{DysonL-K-St-Kopnin}
\end{eqnarray}
See Eq.\ (23) of Ref.\ \onlinecite{Kopnin94}.
In addition, whereas Eq.\ (\ref{DysonL-K}) are given in terms of the
gauge-invariant derivatives of Eqs.\ (\ref{partialT}) and (\ref{partialR}),
terms such as $\mbox{\boldmath$\nabla$}\!{\tilde F}$
and ${\partial}{\tilde F}/{\partial t}$ appear in Kopnin's Eq.\ (21).
It should also be noted that Eq.\ (\ref{DysonL-K}) is free from the
assumption of the slow time variations and is applicable
to the cases of arbitrary external frequencies,
as long as they are much smaller than the Fermi energy.

\section{Self-energy terms}
\label{sec:self-energy}

We next consider the following self-energy terms
appearing on the left-hand side of Eq.\ (\ref{Dyson-R}):
\begin{equation}
\underline{\bar J}(1,2)\equiv 
\,{\rm e}^{iI(\vec{R}_{12},\vec{r}_{1})-iI(\vec{R}_{12},\vec{r}_{2})}
\!\!\int\!\underline{\Sigma}(1,3)\underline{G}(3,2)\, d3 \, ,
\label{SG}
\end{equation}
\begin{equation}
\underline{\bar K}(1,2)
\equiv{\rm e}^{iI(\vec{R}_{12},\vec{r}_{1})-iI(\vec{R}_{12},\vec{r}_{2})}
\!\!\int\!\underline{\Delta}(1,3)\underline{F}^{*\!}(3,2)\, d3 \, ,
\label{DF}
\end{equation}
\begin{equation}
\underline{\bar L}(1,2)
\equiv \,{\rm e}^{iI(\vec{R}_{12},\vec{r}_{1})+iI(\vec{R}_{12},\vec{r}_{2})}
\!\!\int\!\underline{\Sigma}(1,3)\underline{F}(3,2)\, d3 \, ,
\label{SF}
\end{equation}
\begin{equation}
\underline{\bar M}(1,2)
\equiv{\rm e}^{iI(\vec{R}_{12},\vec{r}_{1})+iI(\vec{R}_{12},\vec{r}_{2})}
\!\!\int\!\underline{\Delta}(1,3)\underline{G}^{*\!}(3,2)\, d3 \, .
\label{DG}
\end{equation}
Some of the main issues may be:
(i) whether these terms can also be expressed with respect to the 
gauge-invariant derivatives of Eqs.\ (\ref{partialT}) and (\ref{partialR});
(ii) how the bare mass $m$ in Eq.\ (\ref{DysonL-K})
is changed by the interactions;
(iii) whether new terms arise or not besides the
last four terms in Eq.\ (\ref{DysonL-K}).

We first focus on Eq.\ (\ref{SG}).
Writing it with respect to $\underline{\bar{\Sigma}}$ and $\underline{\bar{G}}$,
it is transformed into an expression
where $\underline{\bar{\Sigma}}\,\underline{\bar{G}}$ is multiplied by a phase factor
${\rm e}^{i\phi_{123}}$ with
\begin{eqnarray}
\phi_{123}\, && \equiv -\frac{e}{c}\oint_{C_{123}}\!\vec{A}(\vec{s})\cdot d\vec{s} 
\nonumber \\
&& = -\frac{e}{2c}\int df^{jk} \left(\frac{\partial A_{k}}{\partial x^{j}}-
\frac{\partial A_{j}}{\partial x^{k}}\right) \, .
\label{phiDef}
\end{eqnarray}
Here the contour $C_{123}$ is given in Fig.\ 1(a), and we have used
the Stokes theorem to obtain the second line,\cite{LL} with
the infinitesimal surface element $df^{jk}$ ($j,k\!=\!0,1,2,3$)
defined by
\begin{figure}[t]
\begin{center}
\leavevmode
\epsfxsize=80mm
\epsfbox{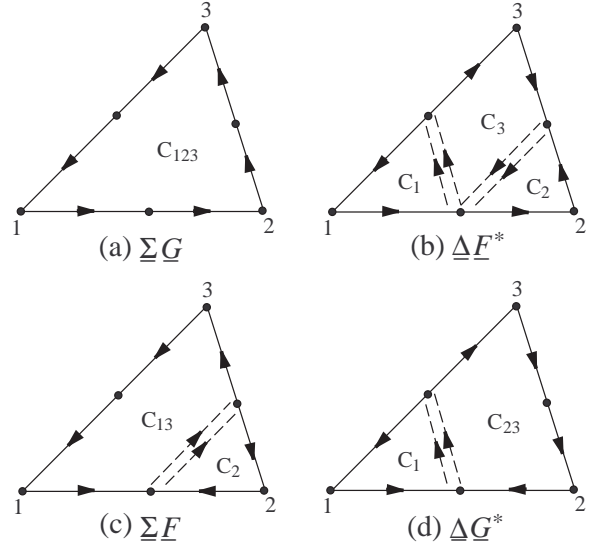}
\end{center}
\caption{Paths of the phase integrals}
\label{fig:1}
\end{figure}
\noindent
\begin{eqnarray}
df^{jk} \equiv && (\vec{r}_{32}du)^{j}(\vec{r}_{13}dv)^{k} 
-(\vec{r}_{32}du)^{k}(\vec{r}_{13}dv)^{j} \, .
\label{df_jk}
\\
&& \hspace{8mm}(0\leq u\leq 1;\hspace{2mm}  0\leq v\leq u)
\nonumber 
\end{eqnarray}
To evaluate Eq.\ (\ref{phiDef}), we expand $B_{jk}\equiv 
\frac{\partial A_{k}}{\partial x^{j}}-
\frac{\partial A_{j}}{\partial x^{k}}$ 
from $\vec{R}\equiv\vec{R}_{12}$ as
\begin{eqnarray}
&& B_{jk}\bigl(\vec{R}+({\textstyle u-\frac{1}{2}})\vec{r}_{32}+
({\textstyle v-\frac{1}{2}})\vec{r}_{13}\bigr) 
\nonumber \\
 = &&
\sum_{n=0}^{\infty}\sum_{n'=0}^{\infty}\frac{(u-\frac{1}{2})^{n}}{n!} 
\frac{(v-\frac{1}{2})^{n'}}{n'!} (\vec{r}_{32}\cdot \vec{\partial}_{\vec{R}})^{n}
(\vec{r}_{13}\cdot \vec{\partial}_{\vec{R}})^{n'}
\nonumber \\
&& \hspace{10mm}\times
B_{jk}(\vec{R}) \, .
\end{eqnarray}
We then find that, with the approximation for $\underline{\Sigma}$
adopted below in Eq.\ (\ref{S-Approx}), 
the terms $(\vec{r}_{13}\cdot \vec{\partial}_{\vec{R}})^{n'}$
with $n'\geq 1$ can be neglected.
This fact may be realized more clearly by noting that
$\vec{r}_{13}\equiv(ct_{13},{\bf r}_{13})$ is transformed in Fourier space into 
$\varepsilon$ and ${\bf p}$ derivatives on 
$\underline{\Sigma}({\bf p}\varepsilon,{\bf R}T)$,
and there already exists one $\vec{r}_{13}$ in the surface element as Eq.\ (\ref{df_jk}).
Using the approximation,
the integrations over $u$ and $v$ in Eq.\ (\ref{phiDef}) are easily performed
to yield
\begin{eqnarray}
\phi_{123}\,&& \approx  -\frac{e}{4} 
\bigl[\, 
2{\cal E}_{1}\!\left({\textstyle \frac{t_{32}}{2}\frac{\partial}{\partial T}}
\right)-{\cal E}_{2}\!\left({\textstyle \frac{t_{32}}{2}\frac{\partial}{\partial T}}
\right)
+{\cal E}_{2}\!\left(-{\textstyle \frac{t_{32}}{2}\frac{\partial}{\partial T}}
\right)\,\bigr]
\nonumber \\
&&\times
\left[ \, \frac{1}{c}{\bf h}(\vec{R})\cdot({\bf r}_{13}\times{\bf r}_{32})
+{\bf E}(\vec{R})\cdot({\bf r}_{13}t_{32}-t_{13}{\bf r}_{32}) \, \right] ,
\nonumber \\
\label{phi123-approx}
\end{eqnarray}
where ${\cal E}_{1}$ and ${\cal E}_{2}$ are defined by
Eqs.\ (\ref{CalE1}) and (\ref{CalE2}), respectively, and
we have neglected spatial derivatives of $\bf E$ and $\bf h$
once again.

We now introduce the Fourier transform of $\underline{\bar{\Sigma}}$ through
\begin{eqnarray}
&&
\underline{\bar{\Sigma}}(1,3)=\int\!\!\frac{d{\bf p}d\varepsilon}{(2\pi)^{4}}
\underline{\Sigma}({\bf p}\varepsilon,{\bf R}_{13}T_{13})
{\rm e}^{i({\bf p}\cdot{\bf r}_{13}-\varepsilon t_{13})}
\nonumber \\
=&&\exp\!\left({\textstyle\frac{{\bf r}_{32}}{2}\!\cdot\!\frac{\partial}{\partial {\bf R}}
\!+\!\frac{t_{32}}{2}\frac{\partial}{\partial T}}\right)\!\!
\int\!\!\frac{d{\bf p}d\varepsilon}{(2\pi)^{4}}
\underline{\Sigma}({\bf p}\varepsilon,{\bf R}T)
{\rm e}^{i({\bf p}\cdot{\bf r}_{13}-\varepsilon t_{13})} \, .
\nonumber \\
\label{S-Fourier}
\end{eqnarray}
Then using Eq.\ (\ref{phi123-approx}),
we can transform Eq.\ (\ref{SG}) into
\begin{eqnarray}
&&\underline{\bar J}=
\int\!\!\frac{d{\bf p}d\varepsilon}{(2\pi)^{4}}\!\!
\int\!\!\frac{d{\bf p}'d\varepsilon'}{(2\pi)^{4}}\!\!
\int\!\!d{\bf r}_{3}dt_{3}\,{\rm e}^{i({\bf p}\cdot{\bf r}_{13}-\varepsilon t_{13})
+i({\bf p}'\cdot{\bf r}_{32}-\varepsilon' t_{32})}
\nonumber \\
&&\times\exp\bigl\{\textstyle \frac{i}{4}e 
\bigl[\, 
2{\cal E}_{1}\!\left(-{\textstyle \frac{i}{2}\partial_{T}\partial_{\varepsilon'}}
\right)-{\cal E}_{2}\!\left(-{\textstyle \frac{i}{2}\partial_{T}\partial_{\varepsilon'}}
\right)
+{\cal E}_{2}\!\left({\textstyle \frac{i}{2}\partial_{T}\partial_{\varepsilon'}}
\right)\,\bigr]
\nonumber \\
&&\hspace{12mm}\times \bigl[ {\textstyle \frac{1}{c}}{\bf h}(\vec{R})\!\cdot\!
(\mbox{\boldmath$\partial$}_{\bf p}\!\times\!\mbox{\boldmath$\partial$}_{{\bf p}'})
\!- \!{\bf E}(\vec{R})\!\cdot\!
(\mbox{\boldmath$\partial$}_{\bf p}\partial_{\varepsilon'}
\!-\!\partial_{\varepsilon}\mbox{\boldmath$\partial$}_{{\bf p}'})\bigr]\bigr\}
\nonumber \\
&& \times\exp\bigl[{\textstyle\frac{i}{2}(
\mbox{\boldmath$\partial$}_{\bf R}\!\cdot\!\mbox{\boldmath$\partial$}_{{\bf p}'}
\!-\!\mbox{\boldmath$\partial$}_{\bf p}\!\cdot\!
\mbox{\boldmath$\partial$}_{{\bf R}'}
\!-\!\partial_{T}\partial_{\varepsilon'}
\!+\!\partial_{\varepsilon}\partial_{T'}})\bigr]
\nonumber \\
&&\times\underline{\Sigma}({\bf p}\varepsilon,{\bf R}T)
\underline{G}({\bf p}'\varepsilon',{\bf R}'T')\bigl|_{{\bf R}'={\bf R},T'=T} \, ,
\label{SG1}
\end{eqnarray}
where $\partial_{T}$ in ${\cal E}_{1}$ and ${\cal E}_{2}$ operates only on ${\bf E}$
and ${\bf h}$.
The latter two exponentials are to be expanded to first order
in $\mbox{\boldmath$\partial$}_{\bf R}$, ${\bf E}$, and 
${\bf h}$, in accordance with Eq.\ (\ref{DysonL-K}).
To this end, we use the following approximation for the self-energy:\cite{Mahan}
\begin{eqnarray}
\underline{\Sigma}({\bf p}\varepsilon,{\bf R}T) \approx
&&
{\rm Re}\underline{\Sigma}({\bf p}_{\rm F}0,{\bf R}T)
+\left(1\!-\!\frac{1}{a}\right)\varepsilon \,\underline{1}
\nonumber \\
&&
\hspace{-15mm}
+\left(\frac{{\bf v}_{\rm F}}{a}-\frac{{\bf p}_{\rm F}}{m}\right)
\cdot({\bf p}\!-\!{\bf p}_{\rm F})\,\underline{1}
+i{\rm Im}\underline{\Sigma}({\bf p}_{\rm F}\varepsilon,{\bf R}T) \, ,
\label{S-Approx}
\end{eqnarray}
with $a$ the renormalization factor and
${\bf v}_{\rm F}$ the Fermi velocity.\cite{comment2}
We then neglect terms including ${\bf R}$ derivatives of 
${\rm Re}\underline{\Sigma}({\bf p}_{\rm F}0,{\bf R}T)$
as well as $\varepsilon$
and ${\bf R}$ derivatives of
${\rm Im}\underline{\Sigma}({\bf p}_{\rm F}\varepsilon,{\bf R}T)$, since
they are smaller at least by an order of magnitude in $(p_{\rm F}\xi_{0})^{-1}$.
We thereby obtain the Fourier transform of Eq.\ (\ref{SG1}) as
\begin{eqnarray}
&&\underline{J}({\bf p}\varepsilon,{\bf R}T)\equiv\int 
\underline{\bar J}(1,2)\,{\rm e}^{-i({\bf p}\cdot{\bf r}-\varepsilon t)}
d{\bf r}dt
\nonumber \\
=&&\underline{\Sigma}({\bf p}\varepsilon,{\bf R}T)\circ
\underline{G}({\bf p}\varepsilon,{\bf R}T)
-\frac{i}{2}
\left(\frac{{\bf v}_{\rm F}}{a}-\frac{{\bf p}_{\rm F}}{m}\right)
\cdot\mbox{\boldmath$\partial$}_{{\bf R}}\underline{G}
\nonumber \\
&&
+\frac{i}{4}e\left(1\!-\!\frac{1}{a}\right)
\bigl[\, 
2{\cal E}_{1}\!\left(-{\textstyle \frac{i}{2}\partial_{T}\partial_{\varepsilon}}
\right)-{\cal E}_{2}\!\left(-{\textstyle \frac{i}{2}\partial_{T}\partial_{\varepsilon}}
\right)
\nonumber \\
&& \hspace{22mm}
+{\cal E}_{2}\!\left({\textstyle \frac{i}{2}\partial_{T}\partial_{\varepsilon}}
\right)\,\bigr]{\bf E}\cdot
\mbox{\boldmath$\partial$}_{\bf p}\underline{G}
\nonumber \\
&&-\frac{i}{4}e 
\bigl[\, 
2{\cal E}_{1}\!\left(-{\textstyle \frac{i}{2}\partial_{T}\partial_{\varepsilon}}
\right)-{\cal E}_{2}\!\left(-{\textstyle \frac{i}{2}\partial_{T}\partial_{\varepsilon}}
\right)
+{\cal E}_{2}\!\left({\textstyle \frac{i}{2}\partial_{T}\partial_{\varepsilon}}
\right)\,\bigr]
\nonumber \\
&&\hspace{5mm}\times
\left(\frac{{\bf v}_{\rm F}}{a}-\frac{{\bf p}_{\rm F}}{m}\right)
\cdot\left(
\frac{1}{c}{\bf h}\!\times\!\mbox{\boldmath$\partial$}_{\bf p}
\!+\!{\bf E}\, \partial_{\varepsilon}\right)\underline{G}
 \, ,
\label{SG2}
\end{eqnarray}
where $A \circ B$ is defined by
\begin{eqnarray}
A\circ B
\equiv && \exp\bigl[{\textstyle\frac{i}{2}(
\partial_{\varepsilon}\partial_{T'}}\!-\!\partial_{T}\partial_{\varepsilon'})
\bigr]
A(\varepsilon T)
B(\varepsilon'T')\bigl|_{
\varepsilon'=\varepsilon,
T'=T} \, .
\nonumber \\
\label{circ}
\end{eqnarray}

We next consider Eq.\ (\ref{DF}).
Writing it with respect to $\underline{\bar{\Delta}}$ and $\underline{\bar{F}}^{*\!}$,
it is expressed as
$\underline{\bar{\Delta}}\,\underline{\bar{F}}^{*\!}$ multiplied by the phase factor:
\begin{eqnarray*}
{\rm e}^{i[I\hspace{-0.2mm}(\hspace{-0.2mm}\vec{R}_{12},\vec{r}_{1})
-I\hspace{-0.2mm}(\hspace{-0.2mm}\vec{R}_{12},\vec{r}_{2})
-I\hspace{-0.2mm}(\hspace{-0.2mm}\vec{R}_{13},\vec{r}_{1})
-I\hspace{-0.2mm}(\hspace{-0.2mm}\vec{R}_{13},\vec{r}_{3})
+I\hspace{-0.2mm}(\hspace{-0.2mm}\vec{R}_{32},\vec{r}_{3})
+I\hspace{-0.2mm}(\hspace{-0.2mm}\vec{R}_{32},\vec{r}_{2})]}\, .
\end{eqnarray*}
We add the four extra paths as the broken lines in Fig.\ 1(b)
and then subtract them.
This factor is thereby written as
\begin{eqnarray*}
&& {\rm e}^{i(\phi_{1}+\phi_{2}+\phi_{3})
+2i[I(\vec{R}_{23},\vec{R}_{12})-I(\vec{R}_{13},\vec{R}_{12})]} 
\, ,
\end{eqnarray*}
where $\phi_{j}$ ($j\!=\! 1,2,3$) is defined as Eq.\ (\ref{phiDef}) with
the contour $C_{j}$ given in Fig.\ 1(b).
Those $\phi_{j}$'s may be transformed into expressions
corresponding to Eq.\ (\ref{phi123-approx}),
but their explicit forms will not be required below;
we should only keep in mind that
they all include $t_{13}$ or ${\bf r}_{13}$,
as in the case of Eq.\ (\ref{phi123-approx}).

We now introduce the Fourier transform of $\underline{\bar{\Delta}}$ 
in the same way as Eq.\ (\ref{S-Fourier}).
We also use the following identity:\cite{Werthamer69}
\begin{eqnarray}
&&\exp[-2iI(\vec{R}_{13},\vec{R}_{12})]
\exp\!\left({\textstyle\frac{{\bf r}_{32}}{2}\!\cdot\!\frac{\partial}{\partial {\bf R}}
\!+\!\frac{t_{32}}{2}\frac{\partial}{\partial T}}\right)
\nonumber \\
= &&\exp\!\left[{\textstyle\frac{{\bf r}_{32}}{2}\!\cdot\!
\left(\frac{\partial}{\partial {\bf R}}\!-\! i\frac{2e}{c}{\bf A}(\vec{R})\right)
\!+\!\frac{t_{32}}{2}\!\left(\frac{\partial}{\partial T}
\!+\! 2ie\Phi(\vec{R})\right)}\right] .
\end{eqnarray}
With these prescriptions, Eq.\ (\ref{DF}) is transformed into
\begin{eqnarray}
&&\underline{\bar K}= 
\int\!\!\frac{d{\bf p}d\varepsilon}{(2\pi)^{4}}\!\!
\int\!\!\frac{d{\bf p}'d\varepsilon'}{(2\pi)^{4}}\!\!
\int\!\!d{\bf r}_{3}dt_{3}\,{\rm e}^{i({\bf p}\cdot{\bf r}_{13}-\varepsilon t_{13})
+i({\bf p}'\cdot{\bf r}_{32}-\varepsilon' t_{32})}
\nonumber \\
&&
\times {\rm e}^{i(\phi_{1}+\phi_{2}+\phi_{3})} \exp\bigl[{\textstyle\frac{i}{2}(
\mbox{\boldmath$\partial$}_{\bf R}\!\cdot\!\mbox{\boldmath$\partial$}_{{\bf p}'}
\!-\!\mbox{\boldmath$\partial$}_{\bf p}\!\cdot\!
\mbox{\boldmath$\partial$}_{{\bf R}'}
\!-\!\partial_{T}\partial_{\varepsilon'}
\!+\!\partial_{\varepsilon}\partial_{T'}})\bigr]
\nonumber \\
&&\times\underline{\Delta}({\bf p}\varepsilon,{\bf R}T)
\underline{F}^{*\!}(-{\bf p}'\!-\!\varepsilon',{\bf R}'T')\bigl|_{
{\bf R}'={\bf R},T'=T} \, ,
\label{DF1}
\end{eqnarray}
where $\partial_{T}$ and $\mbox{\boldmath$\partial$}_{\bf R}$ are defined by
Eqs.\ (\ref{partialT}) and (\ref{partialR}), respectively.
The latter two exponentials should be expanded 
to first order in $\mbox{\boldmath$\partial$}_{\bf R}$, ${\bf E}$, and ${\bf h}$.
To this end, we use the approximation:\cite{comment2}
\begin{eqnarray}
\underline{\Delta}({\bf p}\varepsilon,{\bf R}T) \approx
&&
\underline{\Delta}({\bf p}_{\rm F}0,{\bf R}T) \, .
\label{D-Approx}
\end{eqnarray}
Compared with Eq.\ (\ref{S-Approx}),
the $\varepsilon$ and ${\bf p}-{\bf p}_{\rm F}$ expansions are stopped here
at the lowest level;
this difference originates from the smallness of $(p_{\rm F}\xi_{0})^{-1}$.
With Eq.\ (\ref{D-Approx}), all the ${\bf p}$ and $\varepsilon$ derivatives 
on $\underline{\Delta}$ vanish in Eq.\ (\ref{DF1}), 
so that we may put $\phi_{1}+\phi_{2}+\phi_{3}=0$.
Also, terms including ${\bf R}$ derivatives of
$\underline{\Delta}({\bf p}_{\rm F}0,{\bf R}T)$
should be neglected as they are smaller by an order of magnitude
in $(p_{\rm F}\xi_{0})^{-1}$.
We thereby obtain:
\begin{eqnarray}
&&\underline{K}({\bf p}\varepsilon,{\bf R}T)\equiv\int 
\underline{\bar K}(1,2)\,{\rm e}^{-i({\bf p}\cdot{\bf r}-\varepsilon t)}
d{\bf r}dt
\nonumber \\
=&&\underline{\Delta}({\bf p}_{\rm F}0,{\bf R}T)\circ
\underline{F}^{*\!}(-{\bf p}\!-\!\varepsilon,{\bf R}T) \, .
\label{DF2}
\end{eqnarray}

Equations (\ref{SF}) and (\ref{DG})
are transformed similarly by using the contours of the phase integrals
given in Figs.\ 1(c) and 1(d), respectively.
Especially for Eq.\ (\ref{SF}), we use the approximation:
\begin{eqnarray}
&&\phi_{13}+\phi_{2} \equiv -\frac{e}{c}\oint_{C_{13}}\!\vec{A}(\vec{s})\cdot d\vec{s} 
-\frac{e}{c}\oint_{C_{2}}\!\vec{A}(\vec{s})\cdot d\vec{s} 
\nonumber \\
&& \approx  -\frac{e}{4} 
\bigl[\, 
2{\cal E}_{1}\!\left({\textstyle \frac{t_{32}}{2}\frac{\partial}{\partial T}}
\right)-{\cal E}_{2}\!\left({\textstyle \frac{t_{32}}{2}\frac{\partial}{\partial T}}
\right)
-{\cal E}_{2}\!\left(-{\textstyle \frac{t_{32}}{2}\frac{\partial}{\partial T}}
\right)\,\bigr]
\nonumber \\
&&\hspace{3mm}\times
\left[ \, \frac{1}{c}{\bf h}(\vec{R})\cdot({\bf r}_{13}\times{\bf r}_{32})
+{\bf E}(\vec{R})\cdot({\bf r}_{13}t_{32}-t_{13}{\bf r}_{32}) \, \right] ,
\nonumber \\
\label{phi13+phi2-approx}
\end{eqnarray}
with $C_{13}$ and $C_{2}$ given in Fig.\ 1(c),
which can be derived in the same way as
Eq.\ (\ref{phi123-approx}).
Notice the difference $\pm{\cal E}_{2}
(-{\textstyle \frac{t_{32}}{2}\frac{\partial}{\partial T}})$
between Eqs.\ (\ref{phi123-approx}) and (\ref{phi13+phi2-approx}).
Finally, the Fourier transforms of Eqs.\ (\ref{SF}) and (\ref{DG})
can be written in terms of
the gauge-invariant derivatives of Eqs.\ 
(\ref{partialT}) and (\ref{partialR}) as
\begin{eqnarray}
&&\underline{L}({\bf p}\varepsilon,{\bf R}T)\equiv\int 
\underline{\bar L}(1,2)\,{\rm e}^{-i({\bf p}\cdot{\bf r}-\varepsilon t)}
d{\bf r}dt
\nonumber \\
=&&\underline{\Sigma}({\bf p}\varepsilon,{\bf R}T)\circ
\underline{F}({\bf p}\varepsilon,{\bf R}T)
-\frac{i}{2}
\left(\frac{{\bf v}_{\rm F}}{a}-\frac{{\bf p}_{\rm F}}{m}\right)
\cdot\mbox{\boldmath$\partial$}_{{\bf R}}\underline{F}
\nonumber \\
&&
+\frac{i}{4}e\left(1\!-\!\frac{1}{a}\right)
\bigl[\, 
2{\cal E}_{1}\!\left(-{\textstyle \frac{i}{2}\partial_{T}\partial_{\varepsilon}}
\right)-{\cal E}_{2}\!\left(-{\textstyle \frac{i}{2}\partial_{T}\partial_{\varepsilon}}
\right)
\nonumber \\
&& \hspace{22mm}
-{\cal E}_{2}\!\left({\textstyle \frac{i}{2}\partial_{T}\partial_{\varepsilon}}
\right)\,\bigr]{\bf E}\cdot
\mbox{\boldmath$\partial$}_{\bf p}\underline{F}
\nonumber \\
&&-\frac{i}{4}e 
\bigl[\, 
2{\cal E}_{1}\!\left(-{\textstyle \frac{i}{2}\partial_{T}\partial_{\varepsilon}}
\right)-{\cal E}_{2}\!\left(-{\textstyle \frac{i}{2}\partial_{T}\partial_{\varepsilon}}
\right)
-{\cal E}_{2}\!\left({\textstyle \frac{i}{2}\partial_{T}\partial_{\varepsilon}}
\right)\,\bigr]
\nonumber \\
&&\hspace{5mm}\times
\left(\frac{{\bf v}_{\rm F}}{a}-\frac{{\bf p}_{\rm F}}{m}\right)
\cdot\left(
\frac{1}{c}{\bf h}\!\times\!\mbox{\boldmath$\partial$}_{\bf p}
\!+\!{\bf E}\, \partial_{\varepsilon}\right)\underline{F}
 \, ,
\label{SF2}
\end{eqnarray}
and
\begin{eqnarray}
&&\underline{M}({\bf p}\varepsilon,{\bf R}T)\equiv\int 
\underline{\bar M}(1,2)\,{\rm e}^{-i({\bf p}\cdot{\bf r}-\varepsilon t)}
d{\bf r}dt
\nonumber \\
=&&\underline{\Delta}({\bf p}_{\rm F}0,{\bf R}T)\circ
\underline{G}^{*\!}(-{\bf p}\!-\!\varepsilon,{\bf R}T) \, ,
\label{DG2}
\end{eqnarray}
respectively.

We now collect Eqs.\ (\ref{SG2}), (\ref{DF2}), (\ref{SF2}), 
and (\ref{DG2}) into a Nambu-matrix form
with respect to $\hat{G}({\bf p}\varepsilon ,{\bf R}T)$ and
\begin{eqnarray}
&&\hat{\Sigma}({\bf p}\varepsilon ,{\bf R}T) \equiv \int
\hat{\bar{\Sigma}}(1,2)
\,{\rm e}^{-i({\bf p}\cdot{{\bf r}}-\varepsilon t)} 
\, d{\bf r}dt
\nonumber \\
&& =\left[\!
\begin{array}{cc}
\vspace{1mm}
\underline{\Sigma}({\bf p}\varepsilon ,{\bf R}T) & 
\underline{\Delta}({\bf p}\varepsilon ,{\bf R}T)
\\
-\underline{\Delta}^{\!\! *\!}(-{\bf p}\!-\!\varepsilon ,{\bf R}T) & 
-\underline{\Sigma}^{*\!}(-{\bf p}\!-\!\varepsilon ,{\bf R}T)
\end{array}
\!\right] .
\label{HatSig}
\end{eqnarray}
With the abbreviations 
$\underline{J}\equiv\underline{J}({\bf p}\varepsilon ,{\bf R}T)$
and $\underline{J}^{*}\equiv\underline{J}^{*}(-{\bf p}\!-\!\varepsilon ,{\bf R}T)$,
etc, the third term on the left-hand
side of Eq.\ (\ref{Dyson-R}) can now be written as
\begin{eqnarray}
&&\left[\begin{array}{ll}
-\underline{J}+\underline{K} & -\underline{L}+\underline{M}
\\
-\underline{L}^{*}+\underline{M}^{*}
&
-\underline{J}^{*}+\underline{K}^{*}
\end{array}\right]
\nonumber \\
=&&-\hat{\Sigma}({\bf p}\varepsilon ,{\bf R}T)\circ
\hat{G}({\bf p}\varepsilon ,{\bf R}T)
+\frac{i}{2}
\left(\frac{{\bf v}_{\rm F}}{a}-\frac{{\bf p}_{\rm F}}{m}\right)
\cdot\mbox{\boldmath$\partial$}_{{\bf R}}\hat{\tau}_{3}\hat{G}
\nonumber \\
&&-\frac{i}{4}e\! \left(1\!-\!\frac{1}{a}\right)\!\bigl[ 
2{\cal E}_{1}\!\left({\textstyle -\frac{i}{2}\partial_{T}\partial_{{\varepsilon}}}
\right)\!-\! {\cal E}_{2}\!\left(-{\textstyle \frac{i}{2}\partial_{T}\partial_{{\varepsilon}}}
\right)\bigr]
{\bf E}\cdot \mbox{\boldmath$\partial$}_{{\bf p}}\hat{\tau}_{3}\hat{G}
\nonumber \\
&&-\frac{i}{4}e\!\left(1\!-\!\frac{1}{a}\right)\!
{\cal E}_{2}\!\left({\textstyle \frac{i}{2}\partial_{T}\partial_{{\varepsilon}}}
\right)
{\bf E}\cdot \mbox{\boldmath$\partial$}_{{\bf p}}\hat{G}\hat{\tau}_{3}
\nonumber \\
&&+\frac{i}{4}e\, \bigl[\, 
2{\cal E}_{1}\!\left({\textstyle -\frac{i}{2}\partial_{T}\partial_{{\varepsilon}}}
\right)-{\cal E}_{2}\!\left(-{\textstyle \frac{i}{2}\partial_{T}\partial_{{\varepsilon}}}
\right)\bigr]
\nonumber \\
&&\hspace{20mm}\times
\left(\frac{{\bf v}_{\rm F}}{a}\!-\!\frac{{\bf p}_{\rm F}}{m}\right)
\!\cdot\!
\left(\frac{1}{c}{\bf h}\!\times\!\mbox{\boldmath$\partial$}_{{\bf p}}
+{\bf E}\, \partial_{\varepsilon}\right)\hat{G}
\nonumber \\
&&+\frac{i}{4}e\, {\cal E}_{2}\!\left({\textstyle \frac{i}{2}\partial_{T}\partial_{{\varepsilon}}}
\right)\!
\left(\frac{{\bf v}_{\rm F}}{a}\!-\!\frac{{\bf p}_{\rm F}}{m}\right)
\!\cdot\!
\left(\frac{1}{c}{\bf h}\!\times\!\mbox{\boldmath$\partial$}_{{\bf p}}
+{\bf E}\, \partial_{\varepsilon}\right)\hat{\tau}_{3}\hat{G}\hat{\tau}_{3}
\, ,
\nonumber \\
\label{HatSG}
\end{eqnarray}
with
\begin{eqnarray}
&&\hat{\Sigma}({\bf p}\varepsilon,{\bf R}T)
=\left(1\!-\!\frac{1}{a}\right)\varepsilon \,\hat{1}
+\left(\frac{{\bf v}_{\rm F}}{a}-\frac{{\bf p}_{\rm F}}{m}\right)
\cdot({\bf p}\!-\!{\bf p}_{\rm F})\,\hat{\tau}_{3}
\nonumber \\ 
&&\hspace{5mm}+
\left[
\begin{array}{cc}
{\rm Re}\underline{\Sigma}({\bf p}_{\rm F}0,{\bf R}T) &
\underline{\Delta}({\bf p}_{\rm F}0,{\bf R}T)
\\
-\underline{\Delta}^{\!\! *}\!(-{\bf p}_{\rm F}0,{\bf R}T) &
-{\rm Re}\underline{\Sigma}(-{\bf p}_{\rm F}0,{\bf R}T)
\end{array}
\right]
\nonumber \\
&&\hspace{5mm}+\, i
\left[
\begin{array}{cc}
{\rm Im}\underline{\Sigma}({\bf p}_{\rm F}\varepsilon,{\bf R}T)
&
\underline{0}
\\
\underline{0} &
{\rm Im}\underline{\Sigma}(-{\bf p}_{\rm F}\!-\!\varepsilon,{\bf R}T)
\end{array}
\right]
\, .
\label{SigmaApprox}
\end{eqnarray}
One should keep in mind
that $\partial_{T}$ in ${\cal E}_{1}$ and ${\cal E}_{2}$ 
of Eq.\ (\ref{HatSG}) operates only on
${\bf E}$ and ${\bf h}$.

Several comments are in order before closing the section.
First, Eq.\ (\ref{HatSG}) is expressed entirely in terms of the
gauge-invariant derivatives of Eqs.\ (\ref{partialT}) and (\ref{partialR}).
Second, it exactly contains those terms which turn the bare mass of 
Eq.\ (\ref{DysonL-K}) into the effective mass $m^{*}\equiv 
({\partial v_{\rm F}}/{\partial p_{\rm F}})^{-1}$.
Finally, aside from this change of the bare mass into the effective mass,
no new terms arise in Eq.\ (\ref{HatSG}) besides the last four terms
of Eq.\ (\ref{DysonL-K}).

\section{Quasiclassical equations}
\label{sec:quasi}

Adding Eqs.\ (\ref{DysonL-K}) and (\ref{HatSG}) yields
the Fourier transform of the left-hand side
of Eq.\ (\ref{Dyson-R}).
The corresponding right-hand side is just the unit matrix
$\hat{1}$.
Noting $(\!\varepsilon+\frac{i}{2}\partial_{T})\hat{G}
=\varepsilon\circ\hat{G}$ with $\circ$ defined by Eq.\ (\ref{circ}),
we obtain the left-hand Dyson-Gor'kov equation as 
\begin{eqnarray}
&&(\varepsilon\hat{1}-\hat{\sigma}\hat{\tau}_{3})\circ\hat{G}
-\xi\hat{\tau}_{3}\hat{G}
+\frac{i}{2}{\bf v}_{\rm F}
\cdot\mbox{\boldmath$\partial$}_{{\bf R}}\hat{\tau}_{3}\hat{G}
\nonumber \\
&&+\frac{i}{4}e \bigl[ 
2{\cal E}_{1}\!\left({\textstyle -\frac{i}{2}\partial_{T}\partial_{{\varepsilon}}}
\right)\!-\! {\cal E}_{2}\!\left(-{\textstyle \frac{i}{2}\partial_{T}\partial_{{\varepsilon}}}
\right)\bigr]
{\bf E}\cdot \mbox{\boldmath$\partial$}_{{\bf p}}\hat{\tau}_{3}\hat{G}
\nonumber \\
&&+\frac{i}{4}e
{\cal E}_{2}\!\left({\textstyle \frac{i}{2}\partial_{T}\partial_{{\varepsilon}}}
\right)
{\bf E}\cdot \mbox{\boldmath$\partial$}_{{\bf p}}\hat{G}\hat{\tau}_{3}
\nonumber \\
&&+\frac{i}{4}e\, \bigl[\, 
2{\cal E}_{1}\!\left({\textstyle -\frac{i}{2}\partial_{T}\partial_{{\varepsilon}}}
\right)-{\cal E}_{2}\!\left(-{\textstyle \frac{i}{2}\partial_{T}\partial_{{\varepsilon}}}
\right)\bigr]
\nonumber \\
&&\hspace{20mm}\times
{\bf v}_{\rm F}
\!\cdot\!
\left(\frac{1}{c}{\bf h}\!\times\!\mbox{\boldmath$\partial$}_{{\bf p}}
+{\bf E}\, \partial_{\varepsilon}\right)\hat{G}
\nonumber \\
&&+\frac{i}{4}e\, {\cal E}_{2}\!\left({\textstyle \frac{i}{2}\partial_{T}\partial_{{\varepsilon}}}
\right)\!
{\bf v}_{\rm F}
\!\cdot\!
\left(\frac{1}{c}{\bf h}\!\times\!\mbox{\boldmath$\partial$}_{{\bf p}}
+{\bf E}\, \partial_{\varepsilon}\right)\hat{\tau}_{3}\hat{G}\hat{\tau}_{3}
\nonumber \\
&& = a \hat{1} \, ,
\label{DysonL}
\end{eqnarray}
where $\xi\equiv {\bf v}_{\rm F}\cdot({\bf p}-{\bf p}_{\rm F})$,
and $\hat{\sigma}$ is defined by
\begin{eqnarray}
&&\hat{\sigma}(\hat{\bf p}\varepsilon,{\bf R}T)
\equiv a \left(\frac{p_{\rm F}^{2}}{2m}-\mu\right)\hat{1}
\nonumber \\
&&+
a\left[
\begin{array}{cc}
{\rm Re}\underline{\Sigma}({\bf p}_{\rm F}0,{\bf R}T) &
\underline{\Delta}({\bf p}_{\rm F}0,{\bf R}T)
\\
-\underline{\Delta}^{\!\! *}\!(-{\bf p}_{\rm F}0,{\bf R}T) &
-{\rm Re}\underline{\Sigma}(-{\bf p}_{\rm F}0,{\bf R}T)
\end{array}
\right]\hat{\tau}_{3}
\nonumber \\
&&+i
a\left[
\begin{array}{cc}
{\rm Im}\underline{\Sigma}({\bf p}_{\rm F}\varepsilon,{\bf R}T)
&
\underline{0}
\\
\underline{0} &
{\rm Im}\underline{\Sigma}(-{\bf p}_{\rm F}\!-\!\varepsilon,{\bf R}T)
\end{array}
\right]\hat{\tau}_{3}
\, .
\label{sigmaDef}
\end{eqnarray}
The corresponding right-hand equation may be derived similarly.
It can also be obtained from Eq.\ (\ref{DysonL})
by: (i)  taking its Hermitian conjugate with noting
the relations 
$\hat{G}^{{\rm R}\dagger}({\bf p}\varepsilon,{\bf R}T)
=\hat{G}^{{\rm A}}({\bf p}\varepsilon,{\bf R}T)$ and
$\hat{\Sigma}^{{\rm R}\dagger}({\bf p}\varepsilon,{\bf R}T)
=\hat{\Sigma}^{{\rm A}}({\bf p}\varepsilon,{\bf R}T)$,
where $^{\rm A}$ denotes ``advanced'';
(ii) formally changing $^{\rm A}$ to $^{\rm R}$.
The result is:
\begin{eqnarray}
&&\hat{G}\circ(\varepsilon\hat{1}-\hat{\sigma}\hat{\tau}_{3})
-\xi\hat{G}\hat{\tau}_{3}
-\frac{i}{2}{\bf v}_{\rm F}
\cdot\mbox{\boldmath$\partial$}_{{\bf R}}\hat{G}\hat{\tau}_{3}
\nonumber \\
&&-\frac{i}{4}e \bigl[ 
2{\cal E}_{1}\!\left({\textstyle \frac{i}{2}\partial_{T}\partial_{{\varepsilon}}}
\right)\!-\! {\cal E}_{2}\!\left({\textstyle \frac{i}{2}\partial_{T}\partial_{{\varepsilon}}}
\right)\bigr]
{\bf E}\cdot \mbox{\boldmath$\partial$}_{{\bf p}}\hat{G}\hat{\tau}_{3}
\nonumber \\
&&-\frac{i}{4}e
{\cal E}_{2}\!\left({\textstyle -\frac{i}{2}\partial_{T}\partial_{{\varepsilon}}}
\right)
{\bf E}\cdot \mbox{\boldmath$\partial$}_{{\bf p}}\hat{\tau}_{3}\hat{G}
\nonumber \\
&&-\frac{i}{4}e\, \bigl[\, 
2{\cal E}_{1}\!\left({\textstyle \frac{i}{2}\partial_{T}\partial_{{\varepsilon}}}
\right)-{\cal E}_{2}\!\left({\textstyle \frac{i}{2}\partial_{T}\partial_{{\varepsilon}}}
\right)\bigr]
\nonumber \\
&&\hspace{20mm}\times
{\bf v}_{\rm F}
\!\cdot\!
\left(\frac{1}{c}{\bf h}\!\times\!\mbox{\boldmath$\partial$}_{{\bf p}}
+{\bf E}\, \partial_{\varepsilon}\right)\hat{G}
\nonumber \\
&&-\frac{i}{4}e\, {\cal E}_{2}\!
\left(-{\textstyle \frac{i}{2}\partial_{T}\partial_{{\varepsilon}}}
\right)\!
{\bf v}_{\rm F}
\!\cdot\!
\left(\frac{1}{c}{\bf h}\!\times\!\mbox{\boldmath$\partial$}_{{\bf p}}
+{\bf E}\, \partial_{\varepsilon}\right)\hat{\tau}_{3}\hat{G}\hat{\tau}_{3}
\nonumber \\ 
&& = a \hat{1} \, .
\label{DysonR}
\end{eqnarray}

Let us rewrite the above two equations 
with respect to $\hat{G}' \! \equiv \!\hat{\tau}_{3} \hat{G}$.
We next operate $\hat{\tau}_{3}$ from the left and the right sides
of Eq.\ (\ref{DysonR}),
and subtract the resulting equation from Eq.\ (\ref{DysonL}).
We then perform the integration over $\xi$,
neglecting all the $\xi$ dependences except
that of $\hat{G}'$.
To this end, let us define the quasiclassical Green's function by
\begin{eqnarray}
&&\hat{g}(\hat{\bf p}\varepsilon,{\bf R}T)
\equiv\frac{i}{a\pi}\!\int_{-\infty}^{\infty}
\! \hat{\tau}_{3}\hat{G}({\bf p}\varepsilon,{\bf R}T)\cos(\xi 0_{+}) \, d \xi 
\nonumber \\
&& \hspace{7mm}=\left[\!
\begin{array}{cc}
\vspace{1mm}
\underline{g}(\hat{\bf p}\varepsilon ,{\bf R}T) & 
\underline{f}(\hat{\bf p}\varepsilon ,{\bf R}T)
\\
-\underline{f}^{*\!}(-\hat{\bf p}\!-\!\varepsilon ,{\bf R}T) & 
-\underline{g}^{*\!}(-\hat{\bf p}\!-\!\varepsilon ,{\bf R}T)
\end{array}
\!\right] ,
\label{Hatg}
\end{eqnarray}
with $0_{+}$ an infinitesimal positive constant.\cite{Eilenberger68,Schelankov85}
We also take the following procedures to get the final equations:
(i) Rewrite ${\mbox{\boldmath$\partial$}}_{{\bf p}}\!=\!
{\mbox{\boldmath$\partial$}}_{{\bf p}_{\parallel}}\!+\!
{\bf v}_{\rm F}\frac{\partial}{\partial \xi}$
with ${\bf p}_{\parallel}$ the component on the energy surface $\xi$.
(ii) Notice
${\bf v}_{\rm F}\times{\mbox{\boldmath$\partial$}}_{{\bf p}_{\parallel}}
= {\bf v}_{\rm F}\times{\mbox{\boldmath$\partial$}}_{{\bf p}}$ 
and
\begin{eqnarray*}
\int_{-\infty}^{\infty}
\! d \xi \,  \frac{\partial^{n}}{\partial \xi^{n}}
\hat{\tau}_{3}\hat{G}({\bf p}\varepsilon,{\bf R}T)\cos(\xi 0_{+}) =0 \, .
\end{eqnarray*}
(iii) Neglect terms with
${\bf E}\cdot {\mbox{\boldmath$\partial$}}_{{\bf p}_{\parallel}}$,
since they are smaller than those with
${\bf v}_{\rm F}\cdot{\bf E} \partial_{\varepsilon}$
by an order of magnitude in $(p_{\rm F}\xi_{0})^{-1}$.
(iv) Make use of the integral expressions of Eqs.\ (\ref{CalE1}) and (\ref{CalE2}).

Thus, the quasiclassical equations are obtained as
\begin{eqnarray}
&&[\varepsilon \hat{\tau}_{3}-\hat{\sigma},\hat{g}]_{\circ}
\!+\!i{\bf v}_{\rm F}\cdot\mbox{\boldmath$\partial$}_{\bf R}\hat{g}
\nonumber \\
&&+
\frac{i}{2}{\cal O}_{g}\{\hat{\tau}_{3},\hat{g}\}
+\frac{i}{2}{\cal O}_{f}[\hat{\tau}_{3},\hat{g}]
=\hat{0}
\, ,
\label{QCE}
\end{eqnarray}
where $[A,B]\equiv AB-BA$, $[A,B]_{\circ}\equiv A\circ
B-B\circ A$, 
and ${\cal O}_{g}$ and ${\cal O}_{f}$ are defined by
\begin{eqnarray}
{\cal O}_{g}\equiv 
\frac{1}{2} \int_{-1}^{1}\! d\eta \biggl\{&&
\left[\frac{e}{c}{\bf v}_{\rm F}\!\times\!
{\bf h}\bigl({\bf R},T\!-\!{\textstyle \frac{i}{2}}\eta\partial_{\varepsilon}\bigr)
\right]\cdot
\frac{\partial}{\partial {\bf p}}
\nonumber \\
&&
+ e{\bf v}_{\rm F}\!\cdot\!
{\bf E}\bigl({\bf R},T\!-\!{\textstyle \frac{i}{2}}\eta\partial_{\varepsilon}\bigr)
\frac{\partial}{\partial \varepsilon} \biggr\} \, ,
\label{O1}
\end{eqnarray}
\begin{eqnarray}
{\cal O}_{f}\equiv \,&&
\frac{1}{2} \left(\int_{0}^{1}\!-\!\int_{-1}^{0}\right) \! d\eta \biggl\{
\left[\frac{e}{c}{\bf v}_{\rm F}\!\times\!
{\bf h}\bigl({\bf R},T\!-\!{\textstyle \frac{i}{2}}\eta\partial_{\varepsilon}\bigr)
\right]\cdot
\frac{\partial}{\partial {\bf p}}
\nonumber \\
&&
\hspace{20mm}+ e{\bf v}_{\rm F}\!\cdot\!
{\bf E}\bigl({\bf R},T\!-\!{\textstyle \frac{i}{2}}\eta\partial_{\varepsilon}\bigr)
\frac{\partial}{\partial \varepsilon}\biggr\}  \, ,
\label{O2}
\end{eqnarray}
which operate on $\underline{g}$ and $\underline{f}$ of Eq.\ (\ref{Hatg}), 
respectively.
If $\bf E$ and $\bf h$ are time independent,
these operators acquire the simple expressions:
\begin{eqnarray}
{\cal O}_{g} =
\frac{e}{c}({\bf v}_{\rm F}\!\times\!{\bf h})\cdot
\frac{\partial}{\partial {\bf p}}
+ e{\bf v}_{\rm F}\cdot{\bf E}\frac{\partial}{\partial \varepsilon}
 \, ,
\label{O1T}
\end{eqnarray}
\begin{equation}
{\cal O}_{f} = 0 \, .
\label{O2T}
\end{equation}
Without the terms with ${\cal O}_{g}$ and ${\cal O}_{f}$,
Eq.\ (\ref{QCE}) reduces to the standard quasiclassical equations.
By applying Eq.\ (\ref{QCE}) to the normal state,
one can show easily that Eq.\ 
(\ref{O1T}) indeed describes the normal-state Hall effect.
Thus, Eq.\ (\ref{QCE}) is expected to bring a consistent understanding of the
Hall effect through the superconducting transition.
Notice that the same Fermi velocity ${\bf v}_{\rm F}$ is relevant in both
the acceleration term
$e{\bf v}_{\rm F}\cdot{\bf E}\frac{\partial}{\partial \varepsilon}$
and the 
Lorentz-force term $\frac{e}{c}({\bf v}_{\rm F}\!\times\!{\bf h})\cdot
\frac{\partial}{\partial {\bf p}}$ of Eq.\ (\ref{O1T}),
even after the correlation effects have been incorporated.

A key point in the above derivation is that the terms with 
${\bf E}\cdot {\mbox{\boldmath$\partial$}}_{{\bf p}_{\parallel}}$
turn out to be smaller
than those with
${\bf v}_{\rm F}\cdot{\bf E} \partial_{\varepsilon}$
by an order of magnitude in $(p_{\rm F}\xi_{0})^{-1}$.
Hence those terms in Eqs.\ (\ref{DysonL-K}), (\ref{SG1}), and (\ref{DF1})
could be neglected from the beginning.
Tracing back the derivation,
it then follows that the $\varepsilon$ expansion in Eqs.\ (\ref{S-Approx})
and (\ref{D-Approx}) are unnecessary, so that $\hat{\sigma}$ in Eq.\ (\ref{QCE}) 
may have a more general $\varepsilon$ dependence than Eq.\ (\ref{sigmaDef}) as 
\begin{eqnarray}
&&\hat{\sigma}(\hat{\bf p}\varepsilon,{\bf R}T)
\equiv a \left(\frac{p_{\rm F}^{2}}{2m}-\mu\right)\hat{1} +
a
\hat{\Sigma}({\bf p}_{\rm F}\varepsilon,{\bf R}T)\hat{\tau}_{3}
 \, .
\label{sigmaDef2}
\end{eqnarray}
Hence Eq.\ (\ref{QCE}) can also be used, for example, for a system with a strong electron-phonon
interaction where there may be a strong $\varepsilon$ dependence in $\hat{\Sigma}$.

Equation (\ref{QCE}) carries manifest gauge invariance,
i.e., it remains unchanged in the gauge transformation 
$\hat{g}\!\rightarrow\!\exp[i\frac{e}{c}\chi(\vec{R})\hat{\tau_3}]\hat{g}
\exp[-i\frac{e}{c}\chi(\vec{R})\hat{\tau_3}]$,
$\Phi\!\rightarrow\!\Phi\!-\!\frac{1}{c}\frac{\partial\chi}{\partial T}$, and
${\bf A}\!\rightarrow\!{\bf A}\!+\!\frac{\partial\chi}{\partial {\bf R}}$.
This is certainly a desired property to provide a support for 
the validity of the present equations.
Compared with the result of Kopnin,\cite{Kopnin94} Eq.\ (\ref{QCE}) are more 
advantageous in its wide applicability,
i.e.\ it can be used for clean as well as dirty superconductors 
in arbitrary external frequencies much smaller than the Fermi energy.
In addition, the terms with ${\cal O}_{g}$ and ${\cal O}_{f}$ are also present in 
the retarded and the advanced parts of the equations;
those terms were neglected by Kopnin who considered only the static case of $\omega=0$,
but may have an important role in the vortex dynamics of finite external frequencies.
One can also show that Eq.\ (\ref{QCE}) agrees in the static limit
to the equations obtained by Houghton and Vekhter,\cite{HV98}
if a due care is taken in the gauge choice and terms next order 
in $(p_{\rm F}\xi_{0})^{-1}$ 
(i.e.\ terms with 
${\bf R}$ and ${\bf p}$ derivatives of $\hat{\sigma}$ and 
$\hat{\Delta}$)
are neglected in their Eq.\ (53).
Thus, Eq.\ (\ref{QCE}) clarifies the applicability 
of their Eq.\ (53) that
it is valid only in the static limit;
for example, the first term 
in the square bracket of Eq.\ (\ref{O2}) 
is absent in their equation.

It follows from Eq.\ (\ref{QCE}) that $\hat{\nu}\equiv \hat{g}\circ\hat{g}$
satisfies
\begin{eqnarray}
&&[\varepsilon\hat{\tau}_{3}-\hat{\sigma},
\hat{\nu}]_{\circ}
+i{\bf v}_{\rm F}\cdot \mbox{\boldmath$\partial$}_{\bf R}\hat{\nu}
\nonumber \\
= &&-\frac{i}{2}\{\hat{g},\,
{\cal O}_{g}\{\hat{\tau}_{3},\hat{g}\}\!+\!{\cal O}_{f}
[{\tau}_{3},\hat{g}]\}_{\circ} \, ,
\label{Hatg-Norm}
\end{eqnarray}
with $\{A,B\}_{\circ}\equiv A\circ B+B\circ A$.
In the absence of the right-hand terms, 
this equation tells us that if $\hat{\nu}=\hat{1}$ at some space point,
as in the uniform cases,
then ${\bf v}_{\rm F}\cdot \mbox{\boldmath$\partial$}_{\bf R}\hat{\nu}$
vanishes so that $\hat{\nu}$ does not change
along the straight-line path parallel to ${\bf v}_{\rm F}$;
we may thereby conclude $\hat{\nu}=\hat{1}$ everywhere.
However, this normalization condition no longer holds generally in the
presence of the right-hand terms.
This does not cause any trouble, however, and we only
have to solve Eq.\ (\ref{QCE}) with imposing 
the condition that $\hat{\nu}\rightarrow\hat{1}$ 
as $\varepsilon\rightarrow\infty$ or 
${\bf E}\rightarrow 0$.

\section{Equations in Nambu-Keldysh space}
\label{sec:Nambu-Keldysh}
The above result for the retarded Green's function can easily be 
extended to the advanced and the Keldysh parts.\cite{Rainer83}

Let us define the advanced Green's functions by
\begin{eqnarray}
G_{\alpha\beta}^{\rm A}(1,2) && \, \equiv i\theta(t_{2}\!-\!t_{1})\langle 
\{\psi_{\alpha}(1),\psi_{\beta}^{\dagger} (2)\}\rangle 
\nonumber \\
&& \, = G_{\beta\alpha}^{{\rm R}*}(2,1)\, ,
\label{G-Adef}
\end{eqnarray}
\begin{eqnarray}
F_{\alpha\beta}^{{\rm A}}(1,2) && \, \equiv i\theta(t_{2}\!-\!t_{1})\langle 
\{\psi_{\alpha}(1),\psi_{\beta} (2)\}\rangle
\nonumber \\
&& \,= -F_{\beta\alpha}^{{\rm R}}(2,1) \, .
\label{F-Adef}
\end{eqnarray}
We also define the Fourier transform
and the quasiclassical Green's function
as 
\begin{eqnarray}
&&\hat{G}^{\rm A}({\bf p}\varepsilon ,{\bf R}T) \equiv \int
\hat{\bar{G}}^{\rm A}(1,2) 
\,{\rm e}^{-i({\bf p}\cdot{{\bf r}}-\varepsilon t)} 
\, d{\bf r}dt
\nonumber \\
&& \equiv\left[\!
\begin{array}{cc}
\vspace{1mm}
\underline{G}^{\rm A}({\bf p}\varepsilon ,{\bf R}T) & 
\underline{F}^{\rm A}({\bf p}\varepsilon ,{\bf R}T)
\\
-\underline{F}^{{\rm A}*\!}(-{\bf p}\!-\!\varepsilon ,{\bf R}T) & 
-\underline{G}^{{\rm A}*\!}(-{\bf p}\!-\!\varepsilon ,{\bf R}T)
\end{array}
\!\right] ,
\label{HatGA}
\end{eqnarray}
and
\begin{eqnarray}
&&\hat{g}^{\rm A}(\hat{\bf p}\varepsilon,{\bf R}T)
\equiv\frac{i}{a\pi}\!\int_{-\infty}^{\infty}
\! \hat{\tau}_{3}\hat{G}^{\rm A}({\bf p}\varepsilon,{\bf R}T)\cos(\xi 0_{+}) \, d \xi 
\nonumber \\
&& \hspace{7mm}=\left[\!
\begin{array}{cc}
\vspace{1mm}
\underline{g}^{\rm A}(\hat{\bf p}\varepsilon ,{\bf R}T) & 
\underline{f}^{\rm A}(\hat{\bf p}\varepsilon ,{\bf R}T)
\\
-\underline{f}^{{\rm A}*\!}(-\hat{\bf p}\!-\!\varepsilon ,{\bf R}T) & 
-\underline{g}^{{\rm A}*\!}(-\hat{\bf p}\!-\!\varepsilon ,{\bf R}T)
\end{array}
\!\right] ,
\label{HatgA}
\end{eqnarray}
respectively,
where the arguments $\varepsilon$ and $-\varepsilon$
carry an infinitesimal negative imaginary part.

As for the Keldysh part, we start from the basic definitions:
\begin{equation}
G_{\alpha\beta}^{\rm K}(1,2)  \equiv -i\langle 
[\psi_{\alpha}(1),\psi_{\beta}^{\dagger} (2)]\rangle 
= -G_{\beta\alpha}^{{\rm K}*}(2,1) \, ,
\label{G-Kdef}
\end{equation}
\begin{equation}
F_{\alpha\beta}^{{\rm K}}(1,2)  \equiv -i\langle 
[\psi_{\alpha}(1),\psi_{\beta}(2)]\rangle 
= -F_{\beta\alpha}^{{\rm K}}(2,1) \, .
\label{F-Kdef}
\end{equation}
We then introduce the Nambu matrix by
\begin{eqnarray}
&&\hat{G}^{\rm K}(1,2) \equiv 
\left[\!
\begin{array}{cc}
\vspace{1mm}
\underline{G}^{\rm K}(1,2) & \underline{F}^{\rm K}(1,2)
\\
\underline{F}^{{\rm K}*}(1,2) & \underline{G}^{{\rm K}*}(1,2)
\end{array}
\!\right] .
\label{HatGK-R}
\end{eqnarray}
Its Fourier transform is defined by
\begin{eqnarray}
&&\hat{G}^{\rm K}({\bf p}\varepsilon ,{\bf R}T) \equiv \int
\hat{\bar{G}} ^{\rm K}(1,2) 
\,{\rm e}^{-i({\bf p}\cdot{{\bf r}}-\varepsilon t)} 
\, d{\bf r}dt
\nonumber \\
&& \equiv\left[\!
\begin{array}{cc}
\vspace{1mm}
\underline{G}^{\rm K}({\bf p}\varepsilon ,{\bf R}T) & 
\underline{F}^{\rm K}({\bf p}\varepsilon ,{\bf R}T)
\\
\underline{F}^{{\rm K}*\!}(-{\bf p}\!-\!\varepsilon ,{\bf R}T) & 
\underline{G}^{{\rm K}*\!}(-{\bf p}\!-\!\varepsilon ,{\bf R}T)
\end{array}
\!\right] ,
\label{HatGK}
\end{eqnarray}
and the quasiclassical Green's function by
\begin{eqnarray}
&&\hat{g}^{\rm K}(\hat{\bf p}\varepsilon,{\bf R}T)
\equiv\frac{i}{a\pi}\!\int_{-\infty}^{\infty}
\! \hat{\tau}_{3}\hat{G}^{\rm K}({\bf p}\varepsilon,{\bf R}T)\cos(\xi 0_{+}) \, d \xi 
\nonumber \\
&& \hspace{7mm}=\left[\!
\begin{array}{cc}
\vspace{1mm}
\underline{g}^{\rm K}(\hat{\bf p}\varepsilon ,{\bf R}T) & 
\underline{f}^{\rm K}(\hat{\bf p}\varepsilon ,{\bf R}T)
\\
\underline{f}^{{\rm K}*\!}(-\hat{\bf p}\!-\!\varepsilon ,{\bf R}T) & 
\underline{g}^{{\rm K}*\!}(-\hat{\bf p}\!-\!\varepsilon ,{\bf R}T)
\end{array}
\!\right] .
\label{HatgK}
\end{eqnarray}
The Keldysh self-energy matrices $\hat{\Sigma}^{\rm K}$ and $\hat{\sigma}^{\rm K}$ 
are defined similarly as Eqs.\ (\ref{HatGK}) and (\ref{HatgK}),
respectively.

We now introduce as usual three Keldysh matrices by
\begin{eqnarray}
\check{g}\equiv \!
\left[\!
\begin{array}{cc}
\vspace{1mm}
\hat{g}^{\rm R} & 
\hat{g}^{\rm K}
\\
\hat{0} & 
\hat{g}^{\rm A}
\end{array}
\!\right] , \hspace{2mm}
\check{\sigma}\equiv \!
\left[\!
\begin{array}{cc}
\vspace{1mm}
\hat{\sigma}^{\rm R} & 
\hat{\sigma}^{\rm K}
\\
\hat{0} & 
\hat{\sigma}^{\rm A}
\end{array}
\!\right] ,
\hspace{2mm}
\check{\tau}_{3}\equiv \!
\left[\!
\begin{array}{cc}
\vspace{1mm}
\hat{\tau}_{3} & 
\hat{0} \\
\hat{0} & 
\hat{\tau}_{3}
\end{array}
\!\right] .
\label{Checkg}
\end{eqnarray}
Then the equations for $\hat{g}^{\rm R}$, $\hat{g}^{\rm K}$, and $\hat{g}^{\rm A}$ 
can be put into a compact form as
\begin{eqnarray}
&&[\varepsilon \check{\tau}_{3}-\check{\sigma},\check{g}]_{\circ}
\!+\!i{\bf v}_{\rm F}\cdot\mbox{\boldmath$\partial$}_{\bf R}\check{g}
\nonumber \\
&&+
\frac{i}{2}{\cal O}_{g}\{\check{\tau}_{3},\check{g}\}
+\frac{i}{2}{\cal O}_{f}[\check{\tau}_{3},\check{g}]
=\check{0}
\, ,
\label{QCE2}
\end{eqnarray}
where ${\cal O}_{g}$ and ${\cal O}_{f}$ are defined as
Eqs.\ (\ref{O1}) and (\ref{O2}), respectively.

Those quasiclassical Green's functions satisfy
\begin{equation}
[\hat{g}^{\rm R}(\hat{\bf p}\varepsilon,{\bf R}T)]^{\dagger}=-
\hat{\tau}_{3}\hat{g}^{{\rm A}}(\hat{\bf p}\varepsilon,{\bf R}T)\hat{\tau}_{3} \, ,
\end{equation}
\begin{equation}
[\hat{g}^{\rm K}(\hat{\bf p}\varepsilon,{\bf R}T)]^{\dagger}=\hat{\tau}_{3}
\hat{g}^{{\rm K}}(\hat{\bf p}\varepsilon,{\bf R}T)\hat{\tau}_{3} \, ,
\end{equation}
\begin{equation}
[\hat{g}^{\rm R}(\hat{\bf p}\varepsilon,{\bf R}T)]^{\rm T}=
\hat{\tau}_{2}\hat{g}^{{\rm A}}(-\hat{\bf p}\!-\!\varepsilon,{\bf R}T)\hat{\tau}_{2} \, ,
\end{equation}
\begin{equation}
[\hat{g}^{\rm K}(\hat{\bf p}\varepsilon,{\bf R}T)]^{\rm T}=\hat{\tau}_{2}
\hat{g}^{{\rm K}}(-\hat{\bf p}\!-\!\varepsilon,{\bf R}T)\hat{\tau}_{2} \, ,
\end{equation}
with $^{\rm T}$ denoting the transpose. These relations
originate from $[\hat{G}^{\rm R}({\bf p}\varepsilon,{\bf R}T)]^{\dagger}
=\hat{G}^{\rm A}({\bf p}\varepsilon,{\bf R}T)$, 
$[\hat{G}^{\rm K}({\bf p}\varepsilon,{\bf R}T)]^{\dagger}
=-\hat{G}^{\rm K}({\bf p}\varepsilon,{\bf R}T)$,
$[\hat{G}^{\rm R}({\bf p}\varepsilon,{\bf R}T)]^{\rm T}
=-\hat{\tau}_{1}\hat{G}^{\rm A}(-{\bf p}\!-\!\varepsilon,{\bf R}T)\hat{\tau}_{1}$,
and
$[\hat{G}^{\rm K}(\hat{\bf p}\varepsilon,{\bf R}T)]^{\rm T}
=-\hat{\tau}_{1}\hat{G}^{\rm K}(-{\bf p}\!-\!\varepsilon,{\bf R}T)\hat{\tau}_{1}$,
respectively.
Similarly, the self-energy matrices are shown to have the symmetry:
\begin{equation}
[\hat{\sigma}^{\rm R}(\hat{\bf p}\varepsilon,{\bf R}T)]^{\dagger}=
\hat{\tau}_{3}\hat{\sigma}^{{\rm A}}(\hat{\bf p}\varepsilon,{\bf R}T)\hat{\tau}_{3} \, ,
\end{equation}
\begin{equation}
[\hat{\sigma}^{\rm K}(\hat{\bf p}\varepsilon,{\bf R}T)]^{\dagger}=-\hat{\tau}_{3}
\hat{\sigma}^{{\rm K}}(\hat{\bf p}\varepsilon,{\bf R}T)\hat{\tau}_{3} \, ,
\end{equation}
\begin{equation}
[\hat{\sigma}^{\rm R}(\hat{\bf p}\varepsilon,{\bf R}T)]^{\rm T}=
\hat{\tau}_{2}\hat{\sigma}^{{\rm A}}(-\hat{\bf p}\!-\!\varepsilon,{\bf R}T)\hat{\tau}_{2} \, ,
\end{equation}
\begin{equation}
[\hat{\sigma}^{\rm K}(\hat{\bf p}\varepsilon,{\bf R}T)]^{\rm T}=\hat{\tau}_{2}
\hat{\sigma}^{{\rm K}}(-\hat{\bf p}\!-\!\varepsilon,{\bf R}T)\hat{\tau}_{2} \, .
\end{equation}

\section{Concluding Remarks}
\label{sec:conclusion}
We have presented a systematic derivation of the quasiclassical equations
based a nonlocal-gauge-transformed Green's function (\ref{HatGbar-R}).
This enabled us to retain the gauge invariance
in terms of the center-of-mass coordinate $\vec{R}\!\equiv\!({\bf R}T)$
at every stage throughout the derivation.
Equation (\ref{QCE}) with Eqs.\ (\ref{partialT}), (\ref{partialR}),
(\ref{circ}), (\ref{sigmaDef2}), (\ref{O1}), and (\ref{O2}) 
is the main result of this paper.
It naturally carries a manifest gauge invariance,
i.e., it remains unchanged in the gauge transformation 
$\hat{g}\!\rightarrow\!\exp[i\frac{e}{c}\chi(\vec{R})\hat{\tau_3}]\hat{g}
\exp[-i\frac{e}{c}\chi(\vec{R})\hat{\tau_3}]$,
$\Phi\!\rightarrow\!\Phi\!-\!\frac{1}{c}\frac{\partial\chi}{\partial T}$, and
${\bf A}\!\rightarrow\!{\bf A}\!+\!\frac{\partial\chi}{\partial {\bf R}}$.
This is certainly a desired property to provide a strong support for 
the validity of the present equations.
Also, the terms responsible for the Hall effect 
are automatically present in the operators ${\cal O}_{g}$ and ${\cal O}_{f}$.
Indeed, by applying Eq.\ (\ref{QCE}) to the normal state, one recovers
the normal-state Hall effect.
It should also be noted that Eq.\ (\ref{QCE}) is applicable to band electrons;
this may be shown by using in the derivation
the anisotropic self-energy $\hat{\Sigma}({\bf p}\varepsilon,{\bf R}T)$
where the effect of the periodic lattice potential is incorporated.

Compared with the results of Kopnin\cite{Kopnin94}
and Houghton and Vekhter\cite{HV98} which are valid only in the static limit,
as discussed in the paragraph below Eq.\ (\ref{sigmaDef2}),
Eq.\ (\ref{QCE}) is more advantageous in its wide applicability that
it can be used for clean as well as dirty superconductors up to the 
external frequencies comparable with the energy gap.
In addition, terms with ${\cal O}_{g}$ and ${\cal O}_{f}$ 
are also present in the retarded and the advanced parts of the equations;
those terms were neglected by Kopnin who considered only the static limit
of $\omega=0$, but may have an important role in the cases of finite external frequencies.

Thus, we have derived 
an equation which forms a firm basis for
detailed studies of the Hall effect in the vortex states.
Solving Eq.\ (\ref{QCE}) is expected to bring a comprehensive understanding of
the Hall effect in type-II superconductors.

\acknowledgements

It is a great pleasure to acknowledge extensive and stimulating
discussions on the quasiclassical theory with Dierk Rainer
which led to this work.
I am also grateful to A.\ -P.\ Jauho for an informative communication,
and to the members of 
Physikalisches Institut at Universit\"at Bayreuth
for their hospitality.
The financial support from Yamada Science Foundation
is greatly acknowledged.



\begin{references}
\bibitem{BS65} J. Bardeen and M. J. Stephen,
Phys. Rev. {\bf 140}, A1197 (1965).

\bibitem{NV66}P. Nozi\`eres and W. F. Vinen, Philos. Mag. {\bf 14}, 667 (1966).

\bibitem{Hagen93}For an overview and references, see e.g., 
S. J. Hagen, A. W. Smith, M. Rajeswari,
J. L. Peng, Z. Y. Li, R. L. Greene, S. N. Mao, X. X. Xi,
S. Bhattacharya, Q. Li, and C. J. Lobb, Phys. Rev. B{\bf 47}, 1064 (1993).

\bibitem{Nagaoka98}T. Nagaoka, Y. Matsuda, H. Obara, A. Sawa, T. Terashima,
I. Chong, M. Takano, and M. Suzuki, Phys. Rev. Lett. {\bf 80}, 3594 (1998).

\bibitem{Sonin97}For an overview and references, see e.g.,
E. B. Sonin, Phys. Rev. B{\bf 55}, 485 (1997);
M. Stone, cond-mat/9708017.

\bibitem{Rainer83}For a review, see e.g., J. W. Serene and D. Rainer, 
Phys. Rep. {\bf 101}, 221 (1983); A. I. Larkin and Y. N. Ovchinnikov,
in {\em Nonequilibrium Superconductivity Vol. 12},
ed. by D. N. Langenberg and A. I. Larkin (Elsevier, Amsterdam, 1986) p. 493.

\bibitem{LO95}A. I. Larkin and Y. N. Ovchinnikov,
Phys. Rev.  B{\bf 51}, 5965 (1995).

\bibitem{Kopnin94}N. B. Kopnin, J. Low Temp. Phys. {\bf 97}, 157 (1994).

\bibitem{Kopnin95}N. B. Kopnin and A. V. Lopatin, Phys. Rev. B{\bf 51}, 15291 (1995).

\bibitem{HV98} A. Houghton and I. Vekhter Phys. Rev. B{\bf 57}, 10831 (1998).

\bibitem{Gor'kov59} L. P. Gor'kov, Zh. Eksp. Teor. Fiz. {\bf 36}, 1918 (1959)
[Sov. Phys. JETP {\bf 9}, 1364 (1959)].

\bibitem{Eilenberger66} G. Eilenberger, Z. Phys. {\bf 190}, 142 (1966).

\bibitem{Eilenberger68} G. Eilenberger, Z. Phys. {\bf 214}, 195 (1968).

\bibitem{LO68}A. I. Larkin and Y. N. Ovchinnikov,
Zh. Eksp. Teor. Fiz. {\bf 55}, 2262 (1968)
[Sov. Phys. JETP {\bf 28}, 1200 (1969)].

\bibitem{comment1}
The definitions of the Nambu Green's functions $\hat{G}^{\rm R}$, $\hat{G}^{\rm A}$, 
and $\hat{G}^{\rm K}$ by Eqs.\ (\ref{HatG-R}), (\ref{HatGA}), and (\ref{HatGK-R}), respectively,
are the same as those of the two review articles of Ref.\ \onlinecite{Rainer83}.
On the other hand, the definitions of $\hat{g}^{\rm R}$, $\hat{g}^{\rm A}$, 
and $\hat{g}^{\rm K}$ by Eqs.\ (\ref{Hatg}), (\ref{HatgA}), and (\ref{HatgK}),
respectively,
agree with Larkin and Ovchinnikov but
differ from Serene and Rainer by a factor of ${i}/{\pi}$.


\bibitem{LL} L. D. Landau and E. M. Lifshitz, {\em Classical Theory of Fields}
(Pergamon, Oxford, 1975) 
\S 6.

\bibitem{LF94}M. Levanda and V. Fleurov, J. Phys. Condens. Matter {\bf 6}, 
7889 (1994); 
see also, H. Haug and A. -P. Jauho, 
{\em Quantum Kinetics in Transport and Optics of Semiconductors} 
(Springer-Verlag, Berlin, 1998) Sec.\ 7.

\bibitem{Mahan}See, e.g., G. D. Mahan, 
{\em Many-Particle Physics} (Plenum, NY, 1990)
p. 479-485.

\bibitem{comment2} It turns out eventually that the expansion in terms of
$\varepsilon$ is not required; see the paragraph around Eq.\ (\ref{sigmaDef2}).
However, we here proceed with Eqs.\ (\ref{S-Approx}) and (\ref{D-Approx}),
since they enable us to present a clear derivation.

\bibitem{Werthamer69} N. R. Werthamer, in {\it Superconductivity},
ed. by R. D. Parks (Marcel Dekker, NY, 1969) p. 331.

\bibitem{Schelankov85} A. L. Schelankov, J. Low Temp. Phys. 
{\bf 60}, 29 (1985).

\end{references}
\end{document}